# Quantum-limited optical time transfer for future geosynchronous links

Emily D. Caldwell[1,2,†], Jean-Daniel Deschenes[3,†], Jennifer Ellis[1], William C. Swann[1], Benjamin K. Stuhl[4], Hugo Bergeron[3], Nathan R. Newbury[1,†], and Laura C. Sinclair[1,†]

[1]National Institute of Standards and Technology (NIST), Boulder, CO, USA
[2]Department of Electrical, Energy and Computer Engineering, University of Colorado; Boulder, CO, USA
[3]Octosig Consulting; Quebec City, Quebec, Canada.
[4]Space Dynamics Laboratory; North Logan, Utah, United States.
[†]These authors contributed equally to this work.

**The combination of optical time transfer and optical clocks opens up the possibility of large-scale free-space networks that connect both ground-based optical clocks and future space-based optical clocks. Such networks promise better tests of general relativity[1–3], dark matter searches[4], and gravitational wave detection[5]. The ability to connect optical clocks to a distant satellite could enable space-based very long baseline interferometry (VLBI)[6,7], advanced satellite navigation[8], clock-based geodesy[2,9,10], and thousand-fold improvements in intercontinental time dissemination[11,12]. Thus far, only optical clocks have pushed towards quantum-limited performance[13]. In contrast, optical time transfer has not operated at the analogous quantum limit set by the number of received photons. Here, we demonstrate time transfer with near quantum-limited acquisition and timing at 10,000 times lower received power than previous approaches[14–24]. Over 300 km between mountaintops in Hawaii with launched powers as low as 40 µW, distant timescales are synchronized to 320 attoseconds. This nearly quantum-limited operation is critical for long-distance free-space links where photons are few and amplification costly -- at 4.0 mW transmit power, this approach can support 102 dB link loss, more than sufficient for future time transfer to geosynchronous orbits.**

Comb-based optical time transfer (OTT) follows previous microwave two-way time-frequency transfer[25]. Optical pulses from coherent frequency combs located at remote sites are exchanged across a two-way free-space link. The difference in the detected pulse time-of-arrival between sites yields their clock offset, independent of the time-of-flight (assuming full reciprocality). Previous comb-based OTT used linear optical sampling (LOS) against a local frequency comb with an offset repetition rate to scan across the incoming comb pulses and measure their timing[14–24]. This approach is photon inefficient and requires signals of a few nanowatts, 40 dB above the quantum limit. Despite this, with a combination of 40-cm aperture telescopes, adaptive optics, and watt-level amplifiers, Shen *et al.* (24) achieved a working range of 113 km. The alternative approach of conventional optical frequency transfer using continuous wave (CW) lasers achieves high performance[26–28] but is unable to measure the elapsed time between sites - the quantity of interest to most applications - in the presence of link disruption due to atmospheric turbulence, weather or multiplexed operation.

In close analogy with optical clocks, the quantum-limited uncertainty for time transfer via an optical pulse of width $\tau_p$ (here ~350 fs) is simply

$$\sigma_t = \gamma \frac{\tau_p}{\sqrt{n}} \qquad (1)$$

where $n$ is the number of detected photons in the measurement interval and $\gamma$ is a constant of order unity. Here, we demonstrate OTT at this quantum limit by exploiting the precision and agility of a time programmable frequency comb (TPFC)[29] in conjunction with Kalman filter-based signal processing. The improvement over previous LOS-based OTT is large: the minimum received power decreases 10,000-fold from a few nanowatts to a few hundred femtowatts, or a mean photon number of 0.01 per frequency comb pulse.

We demonstrate this quantum-limited OTT by synchronizing two optical timescales across two different free-space links: a 2-km link with low turbulence in Boulder, CO and a 300-km link with strong turbulence between two mountaintops in Hawaii. Under low turbulence, where the free-space path is indeed reciprocal, the two-way time transfer is nearly quantum-limited; the clocks are synchronized to $246 \text{ as}/\sqrt{P\tau}$ in time and to $4.3\times10^{-16}/\sqrt{P\tau^3}$ in fractional frequency, where $P$ is the received power in picowatts and $\tau$ is the averaging time in seconds, with respective floors of ~35 attoseconds and below $10^{-18}$. Over the 300-km horizontal link, the one-way timing signals are still measured at nearly the quantum limit with a power threshold of 270 fW. However, the strong integrated turbulence leads to excess non-reciprocal time-of-flight noise, attributed to multipath effects. Nevertheless, the clocks remain synchronized to 320 attoseconds in time and $3.1\times10^{-19}$ in frequency at longer averaging times. Finally, synchronization is achieved even at attenuated comb powers of 40 μW with a median received power of 150 fW. This low-power performance is enabled by the robust Kalman-filter approach which can tolerate the >70% signal fades. In comparison to the longest previously reported range for LOS-based OTT[24], quantum-limited OTT operates across 3× the distance, at 20× improved update rate using 200× less comb power, and at 4× lower aperture diameter, with over 14 dB greater tolerable link loss. Importantly, the tolerable link loss of 102 dB exceeds that of future ground to GEO links with similar 10-cm apertures and milliwatt comb powers.

**Quantum-Limited Time Transfer Using a Time Programmable Frequency Comb**

The quantum-limited OTT was demonstrated first over a 2-km link at the NIST campus in Boulder[15,18] and then over the 300-km link between the Hawaiian Islands shown in Figure 1a. For both, a folded link geometry enabled direct out-of-loop verification of the synchronization, at the cost of added link loss, but the system could ultimately be used in a point-to-point[21,24] or multi-node geometry[20]. For the Hawaii link, operation was mainly limited to overnight hours because of daytime clouds in the inter-island convergence zone, but the system operates equally well in daylight and full darkness.

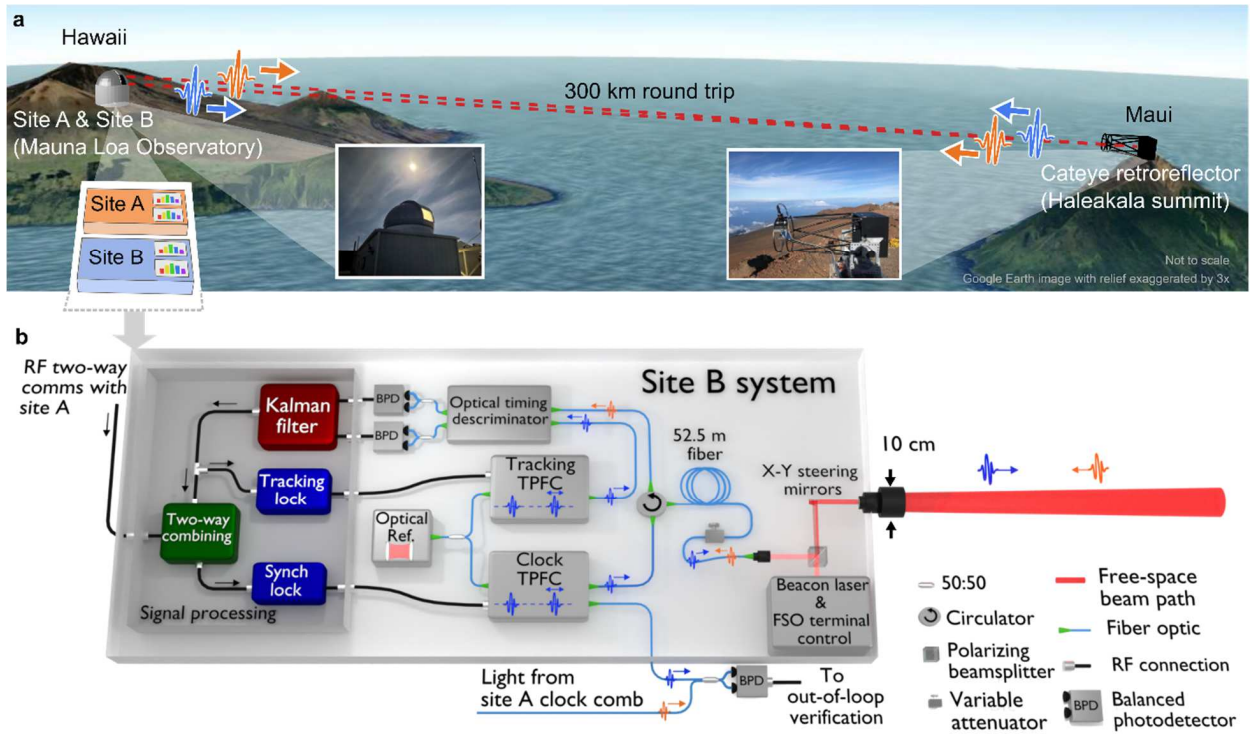

**Figure 1: Quantum-limited optical time transfer. (a) The system was tested on a 300-km folded free-space link established between the two sites, co-located at the Mauna Loa Observatory, and a cateye retroflector located on the summit of Haleakala. Co-location of the sites allows a direct, out-of-loop timing verification. (b) Detailed schematic for site B. (Site A is identical except without the synchronization (Synch) lock.) Each site uses two time-programmable frequency combs (TPFCs) phase locked to a local optical reference. The clock TPFC output defines the local timescale. It is also transmitted across the bi-directional link through a 10-cm aperture free-space optical (FSO) terminal (see Supplemental Figure 1). The tracking TPFC is used to acquire, track, and measure the timing of that incoming clock comb pulse train. Two-way combination of the measured timing signals generates an error signal that is applied at site B for synchronization (see text). The filtered comb output powers are 4.0 mW and 5.9 mW for site A and B (see Supplemental Figure 2), but can be attenuated by the in-line attenuator to mimic links with higher loss. The fiber spools before the FSO terminals compensate for the 300 km of air dispersion.**

The system is centered around fiber-based, 200-MHz repetition frequency, time programmable frequency combs (TPFC) that provide real-time attosecond-level digital control of the pulse timing. A heterodyne timing discriminator[29] measures the time offset between a local tracking TPFC and the incoming clock comb pulse signal with shot-noise limited sensitivity (Supplemental Figure 3). This time offset acts as an error signal to adjust the digital control of the tracking TPFC to follow the incoming clock comb pulses. (See Methods and Supplemental Figures 4-5) The commanded tracking-TPFC timing then replicates the timing of the incoming comb pulse train at each site, $t_A$ or $t_B$, whose difference,

$$\Delta t = \frac{t_A - t_B}{2} = \left(\Delta T_{osc} - \Delta T_{cntrl}\right) + \varepsilon_{NR,turb} + \varepsilon_{qn} + \varepsilon_{combs} \quad (2)$$

is a measure of the time offset between the two clock combs, $\left(\Delta T_{osc} - \Delta T_{cntrl}\right)$, where $\Delta T_{osc}$ is the time offset between the local reference oscillators and $\Delta T_{cntrl}$ is the synchronization feedback to the clock TPFC at site B ($\Delta T_{cntrl} = 0$ for open-loop operation). The fundamental reciprocity of a single spatial mode link[30] means that the time-of-flight, including turbulence effects, should cancel in this two-way comparison of Eq. (2). Nevertheless, we include a non-reciprocal, turbulence noise term, $\varepsilon_{NR,turb}$, for reasons discussed later. The quantum noise term, $\varepsilon_{qn}$, has a standard deviation following Eq. (1). The system noise, $\varepsilon_{combs}$, is typically negligible at short averaging times and low powers, but leads to the flicker floor at long averaging times.

To track the incoming pulses, the incoming clock comb and local tracking comb pulses must overlap in time to within the $\sim 2\tau_p$ dynamic range of the timing discriminator. This initial alignment is accomplished by sweeping the tracking comb pulse position in time[29] and searching for peaks in the heterodyne signal (Figure 2a). Scintillation from turbulence, however, causes random 100% intensity fluctuations which complicate mapping the peak heterodyne voltage to the incoming pulse location. By use of a Kalman filter to aggregate intermittent observations of pulse overlap, we can track the estimated temporal position of the incoming pulses, and the associated uncertainty, despite fades. As the estimated position uncertainty decreases, the search space narrows. When the estimated uncertainty reaches 500 fs the tracking lock is engaged.

After acquisition, $t_A$ and $t_B$ are measured at 40 kSa/s with a 26 kHz bandwidth from the timing discriminator (see Methods). For robust operation through signal fades, these timing samples are input into the Kalman filter to generate optimal estimates of the timing with 10-25 Hz effective bandwidth. These Kalman-filtered values are used in Eq. (2) and input to a 15-Hz bandwidth synchronization lock to steer the site B clock comb. Here, the timing signals used in the two-way combining are communicated from sites A to B by coaxial cable but an optical communications link could be implemented as in Ref. 15. The system is calibrated so the pulses from the two clock combs overlap at the out-of-loop verification reference plane when $\Delta t \to 0$ (Supplemental Figure 6).

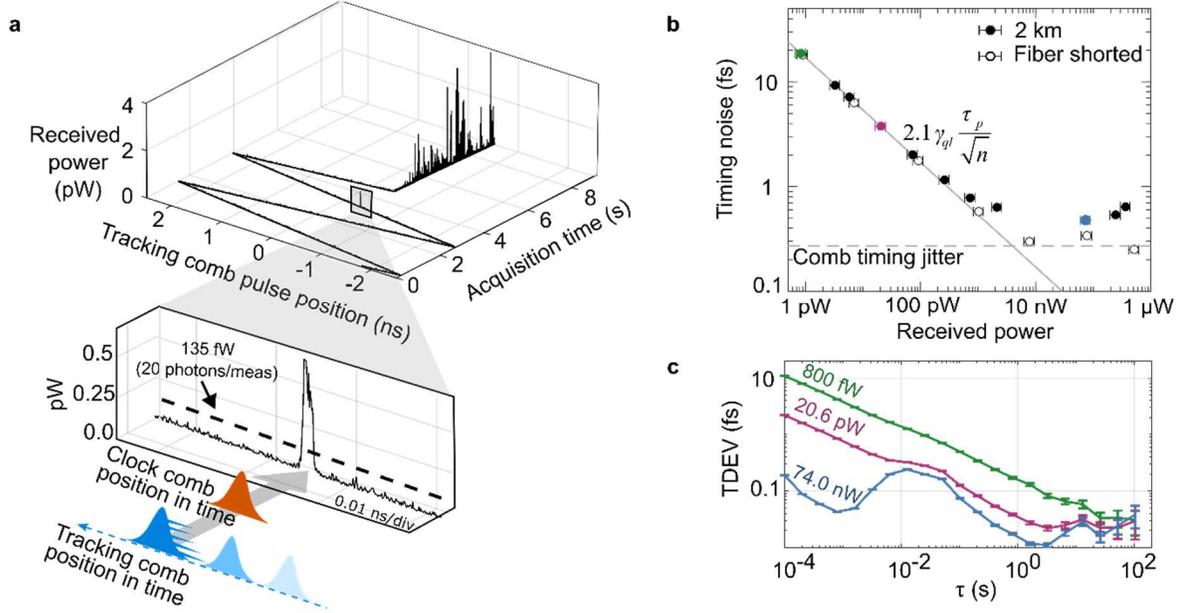

**Figure 2: Low-power acquisition and quantum-limited performance.** (a) Demonstrated signal acquisition over the 300-km Hawaii link. Initially, the local tracking TPFC is swept over its full 5-ns non-ambiguity range in a triangular waveform. At ~3 seconds into the acquisition, a peak in the heterodyne signal indicates a transient temporal overlap between the tracking TPFC and the incoming clock comb (see inset). Based on the observation of a heterodyne signal above the 135 fW threshold, the signal processor steers the tracking comb back to this location for finer search before initiating the tracking lock at ~5 seconds into the acquisition. (b) The timing noise (standard deviation over 600 seconds) in $\Delta t$ measured over a shorted link (open circles) and a 2-km free-space link (closed circles). Colors correspond to traces in (c) below. The timing follows the expected quantum-limit from Eq. (1) for $\gamma = 2.1\gamma_{ql}$ (grey line), where $\gamma_{ql} \approx 0.6$ is the quantum limit for gaussian pulses (see Ref. 29 and Supplemental Table 1). (c) Time deviations (TDEV) over the 2-km free-space link at received powers of 800 fW and 20.6 pW, which follow the quantum limit, until reaching the system noise floor at received powers above 10 nW.

At low power and weak turbulence, the two-way time transfer is nearly quantum-limited following Eq. (1) until it reaches the system noise floor (Figure 2b-c). We apply a ~270-fW threshold on the received power for a valid timing measurement, chosen such that the quantum-limited timing noise standard deviation was approximately one-sixth the full timing discriminator dynamic range (see Supplemental). For comparison, the power threshold for signal acquisition is lower, at ~135 fW, selected to limit false detections to <1/day (see Methods). These thresholds correspond to $n=40$ and 20 photons per 19-μs timing interval, respectively, or 0.01 and 0.005 mean photons per comb pulse. The values of $n>1$ reflect the conservatively chosen threshold to ensure low probability of false detection. Both the acquisition and timing measurements operate at ~2× the quantum limit, due to detector noise power penalty and differential chirp between the tracking comb and incoming comb pulses.

**Demonstration Over a 300-km Terrestrial Free-Space Link**

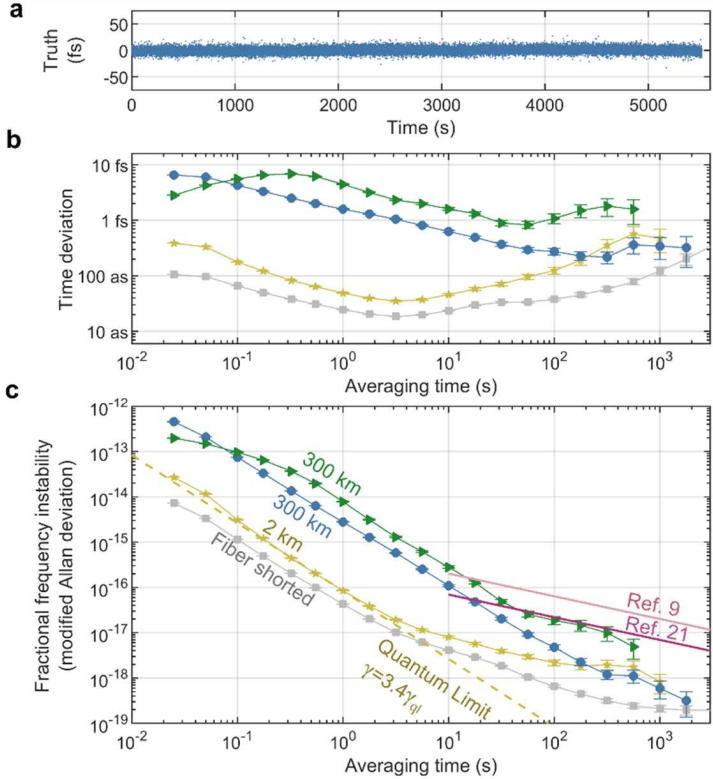

Figure 3: Optical time transfer results measured by the out-of-loop timing comparison. (a) Synchronized clock time difference across the 300-km link for 4.0-mW comb power and 14-pW median received power. (b) and (c) Time deviation and fractional frequency instability (modified Allan deviation) across the 300-km link at 4.0-mW comb power with 14-pW median received power (blue circles) and 40-µW attenuated comb power with 150-fW median received power (green triangles). Also shown are data from the 2-km link at 30-pW received power (yellow stars), and shorted (grey dots). At low turbulence over the 2-km link, the performance is nearly quantum limited (yellow dashed line for $\gamma = 3.4\gamma_{ql}$). Over 300 km, performance is limited by non-reciprocal multipath atmospheric turbulence at short times. The time deviations follow $1.6\,\text{fs}/\sqrt{\tau}$ and $49\,\text{as}/\sqrt{\tau}$ for the 300-km and 2-km links, plateauing at 475 as. The corresponding modified Allan deviations follow $2.8\times10^{-15}\tau^{-3/2}$ and $8.5\times10^{-17}\tau^{-3/2}$, reaching a floor of $3.1\times10^{-19}$. For context, instability curves are provided for comparisons of physically separated clocks involving transportable[9] (light pink line) and laboratory optical clocks[21] (dark pink line).

Figure 3 compares the performance over the 300-km and 2-km links in terms of timing synchronization, timing instability and frequency instability. Additional data is provided in the Supplemental Figures 7-10. Although the one-way timing measurements are quantum limited over the strongly turbulent 300-km link, unlike the shorted and 2-km link, their two-way subtraction, $\Delta t$, does not reach the quantum limit for reasons discussed below. Nevertheless, it drops below state-of-the-art transportable optical atomic clocks[9] at only 6 seconds of averaging time and laboratory optical atomic clocks at 17 seconds of averaging time[21].

To demonstrate operation at extreme link loss, the comb power from site B was attenuated to 40 µW leading to a median received power of 150 fW at site A. Despite signal fading of 73% time below threshold, timing acquisition and synchronization were still achieved with minimal 2.8× performance degradation at short times. This attenuation of the site B power is done in the two-way path and is equivalent to operation with 4.0 mW of comb power over a total link loss of 106 dB.

**Impacts of Turbulence on Timing**

The increased timing noise across the 300-km link is attributed to the strong integrated turbulence and a breakdown in the expected reciprocity in time-of-flight over the single mode link[30]. Previous comb-based OTT has not seen clear violations in reciprocity even at 100 km[14–24]. However, the enhanced sensitivity of quantum-limited OTT means we can probe timing fluctuations at the attosecond-level during deep signal fades when the effects of multipath interference are at their strongest.

This excess noise is illustrated best in the power spectral densities (PSDs) of $\Delta t$ and its counterpart (Figure 4). The latter measures the time-of-flight and shows the expected piston noise (pulse-to-pulse) timing jitter with its $f^{-8/3}$ Kolmogorov scaling[31]. Differential piston noise associated with the 1-ms time-of-flight is negligible, although significant for longer links, as discussed below. Note that the use of a folded link geometry only increases, rather than cancels, the piston-induced time-of-flight noise.

The PSD of $\bar{t}$ exhibit excess time-of-flight noise (shaded region), beyond the piston noise, out to the Greenwood frequency of ~300 Hz before dropping the quantum-limited white noise floor. The PSD of $\Delta t$ (open loop) follows the expected $f^{-3}$ phase noise of the reference oscillators but shows similar excess timing noise (shaded purple region), although suppressed by ~7-11 dB from $\bar{t}$. This excess noise on $\Delta t$ limits the out-of-loop synchronization, yielding the white noise floor shown in Figure 4 and the corresponding elevated instabilities in Figure 3. We attribute it to multi-path interference due to strong atmospheric turbulence across the 300-km horizontal link. Multi-path interference causes pulse distortions that depend on the pulse chirp, which differs between clock combs and is therefore not fully reciprocal leading to the limited 7-11 dB suppression. This pulse distortion has been analyzed in terms of the resulting average pulse temporal broadening, yielding an estimate of the excess $\delta t_{excess} \approx 1.8 L_0^{5/6} c^{-1} \left( L C_n^2 \right)^{-1/2}$ [31,32], where $L_0$ is the outer scale, $c$ is the speed of light, $C_n^2$ is the turbulence structure function and $L$ is the link distance. The inset of Figure 4 shows good agreement between the measured excess and the estimate assuming $L_0 = 1 \text{ m}$, the measured reciprocal suppression, and an integrated turbulence strength, $LC_n^2$, based on a fit to the piston noise for 10 m/s wind speed.

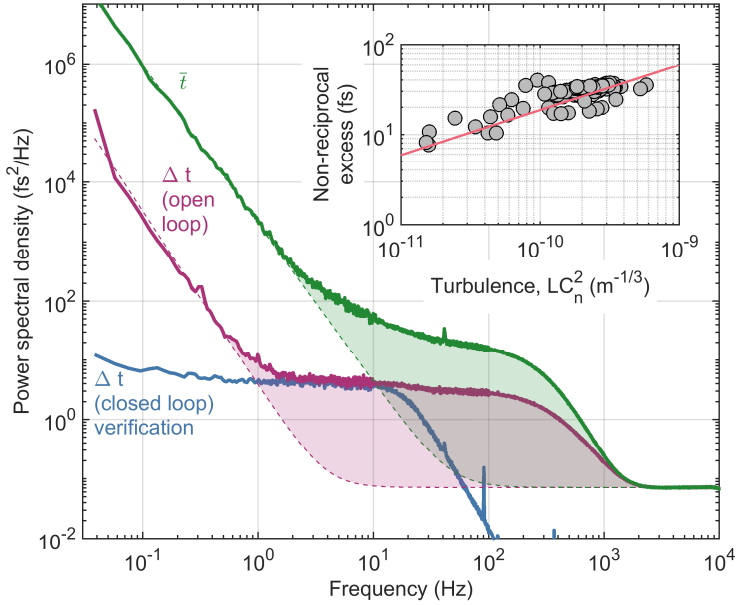

**Figure 4: Power spectral densities measured over the strong turbulence of the 300-km link. Timing power spectral density data for a representative 90-minute measurement for $\Delta t$ (purple solid line), $\bar{t}$ (green solid line), and the out-of-loop synchronization measurement (blue solid line). The shaded regions indicate excess noise beyond the expected reference oscillator noise and piston noise values, for $\Delta t$ and $\bar{t}$ respectively, as well as the shared quantum noise floor (dashed lines). The excess timing noise on $\Delta t$ sets the noise floor for the out-of-loop timing below the synchronization bandwidth of 15 Hz. Inset: Non-reciprocal excess versus $LC_n^2$ for 10-minute intervals over many runs (grey circles). The solid red line is the simple model discussed in the text.**

## Implications for Future Long-distance Ground-to-Space Links

Figure 5 puts the performance and size-weight-power-and-cost (SWaP-C) of quantum-limited OTT in context with previous work and future space-based OTT. The current 10-cm apertures and 4.0 mW comb powers bode well for future low-SWaP-C space instruments[33]. Moreover, the use of 10-cm apertures for the ground as well as satellite terminals permits compact ground stations. For applications such as global time distribution, clock-based geodesy or VLBI, the satellite would include OTT terminals and a high-performance oscillator. Relativistic tests or dark matter searches would require an atomic optical clock onboard the satellite, along with the OTT terminals.

While GEO-to-ground links are 100× farther than the Hawaii link, the path integrated turbulence is ~100× lower. As a result, for paired 10-cm aperture telescopes, the total link loss is similar as shown in Figure 5 (see Supplemental) and the conservative 102 dB tolerable loss exceeds the estimated ground-to-GEO link loss by 11 dB. A fourfold increase in the ground aperture should increase the reach to cislunar orbits. Because the integrated turbulence is about 100× lower than for the Hawaii link, ground-to-space OTT should not suffer from the same level of non-reciprocal timing noise from multi-path effects. However, the turbulence-induced piston noise will no longer perfectly cancel since the up-going and down-going comb pulses traverse the turbulent atmosphere with a 0.12-s time offset, approximately the time-of-flight to GEO. As a result, there is a differential piston noise contribution that limits the time deviation to $2.3 \text{ fs}/\sqrt{\tau}$ and modified Allan deviation to $3.9\times10^{-15}\tau^{-3/2}$ (see Supplemental). Purely coincidentally, these values are close to the data of Figure 3 although of very different origin. Nevertheless, the residual instability is below state-of-the-art optical clocks for $\tau > 10$'s of seconds. OTT could then compare two optical clocks in different locations without degradation via common view, or a ground clock to a space-based clock for tests of relativity[1–3]. Active clock synchronization would be limited to a bandwidth of ~1/(8×0.12 s) =1 Hz, or ~10× lower than the Hawaii link. Therefore, applications that require high coherence between ground and space might combine active synchronization (for observation times beyond 2 s) with near-real-time post-processing of the time samples (for observation times below 2 s).

Although the link margin is more than sufficient for OTT to low or mid-Earth orbits, these involve up to 8 km/s orbital velocities with significant resulting Doppler shifts and point-ahead effects. The latter have been shown to be negligible after correction[34–36]. The Doppler shifts must be known to within the 26-kHz detection bandwidth used here, corresponding to a closing velocity uncertainty of ±2 cm/s, which requires more precise orbit determination than for a GEO orbit. Lacking that, more advanced signal acquisition would be needed, possibly based on a hybrid approach that combines comb-based OTT with CW-laser-based optical frequency transfer.

Figure 5 includes several optical communication missions. Quantum-limited OTT might piggyback on future coherent optical communication links, which would be facilitated by its low power threshold of −96 dBm either by use of a shared aperture or separately steered sub-aperture. It also includes several demonstrations of optical frequency transfer (OFT) via two-way transmission of a CW laser, which reaches extremely low instabilities over low turbulence, km-scale links[26–28]. However, signal interruptions from turbulence, weather, platform motion, and/or multiplexed operation to different sites will result in penalties in the frequency transfer[37] and, critically, prevent measurement of the relative time offset between clocks, the quantity of interest for almost all applications. Quantum-limited OTT, on the other hand, can track the relative time with femtosecond precision even if the link availability is limited to seconds interspersed across hours, days or longer.

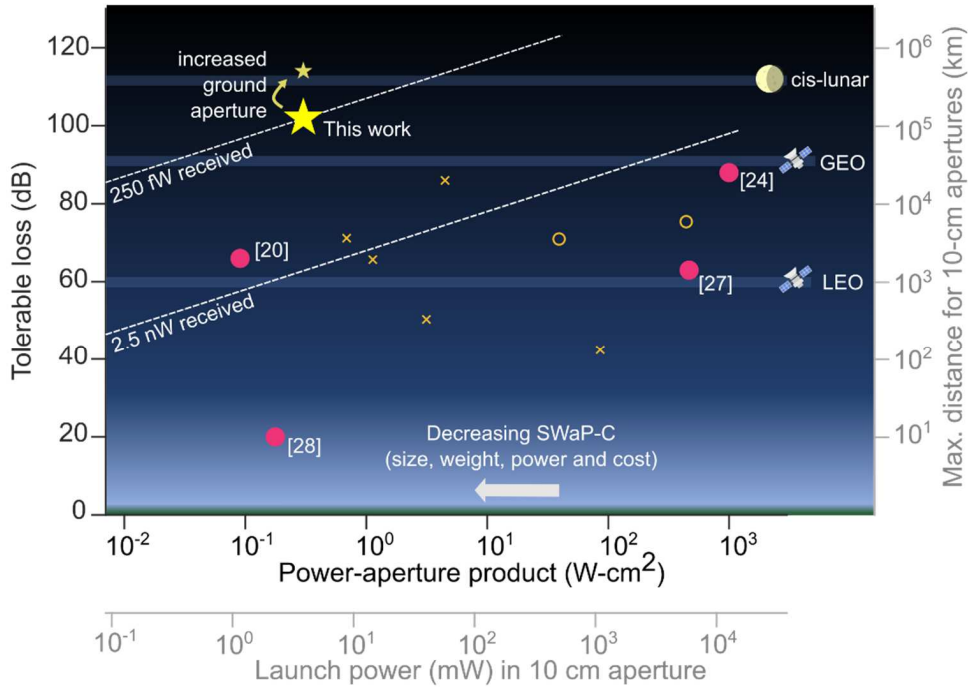

**Figure 5: Ground-to-space time transfer. Future free-space optical time transfer will require operation at long distance, set by tolerable link loss, and low SWaP-C, largely set by the power-aperture product. The current demonstration (large yellow star) achieves the highest tolerable loss of 102 dB at the very low power-aperture product of 0.3 W-cm$^2$ as compared to previous comb-based time-frequency transfer or optical frequency transfer (magenta circles)[20,24,27,28]. The linear tradeoff (white dashed lines) between tolerable loss/maximum distance and power-aperture product is set by the received power threshold, as shown for the LOS-comb based OTT (2.5 nW) and quantum-limited OTT (270 fW) with both curves assuming a 10-cm aperture. For an additional comparison, several optical communication missions are shown (orange circles for GEO and orange x's for LEO)[38–43] The lower bottom and far right axes assume matched 10-cm apertures, 2-dB channel loss, 6-dB total transceiver loss and 6-dB additional coupling loss. (See Supplemental). The maximum projected distance for the current OTT exceeds the 35,786-km GEO altitude and even reaches the 10× farther cislunar distances if the ground aperture is increased to 40 cm, while leaving the space-based aperture at 10 cm (small yellow star).**

**Conclusion**

By operating near the quantum limit, this OTT requires 10,000× less power than previous LOS approaches. Here, we show attosecond time transfer over record distances of 300 km and record link losses of 102 dB, while transmitting only the 4.0 mW output from a compact, unamplified frequency comb. At low turbulence strengths, the time deviation is as low as 35 attoseconds at 3 seconds averaging. Under strong turbulence, the time deviation drops as low as 220 attoseconds at ~300 seconds of averaging, well below what is needed for distributed coherent sensing, redefinition of the second, and tests of fundamental physics[1–12].

The combination of this level of performance, low power-aperture product and the ability to operate at 102 dB link loss without adaptive optics will enable future low SWaP-C OTT from geosynchronous orbits (GEO) to portable ground-based clocks for time transfer, geodesy, relativity tests or distributed sensing. Indeed, with a 4× increase in ground terminal aperture, future OTT could even reach cislunar orbits.

**ACKNOWLEDGEMENTS**

We acknowledge comments from Tobias Bothwell, Fabrizio Giorgetta and Brian Washburn. We acknowledge technical assistance from Saad Syed, Martha Bodine, Holly Leopardi, Tyler Wright, Matthew Martinsen, Darryl Kuniyuki, Sean Baumann, the NOAA Mauna Loa Observatory, and the Haleakala MEES Observatory. Approval no. AFRL-2022-5993.

**FUNDING**
Air Force Office of Scientific Research (MIPR F4FGA02152G001)
OSD and DARPA DSO through a CRADA with Vector Atomic,
National Institute of Standards and Technology (NIST)
Air Force Research Laboratory (FA9453-16-D-0004).

**AUTHOR CONTRIBUTIONS**
JDD, LCS, NRN conceived of the experiment. EDC, JDD, LCS, NRN acquired and analyzed the synchronization data from Hawaii, and wrote the paper. JDD, HB developed the signal acquisition processing. JE, EDC, LCS constructed the optical system and acquired 2-km data. BKS and WCS designed and built the free-space optical terminals, and contributed to the writing.

**MATERIALS AND CORRESPONDANCE**
Should be addressed to Laura Sinclair (laura.sinclair@nist.gov) or Nathan Newbury (nathan.newbury@nist.gov).

# METHODS

## 1. Optical Time Transfer System Design

Supplemental Figure 3 shows the optical design of the OTT transceiver while Supplemental Figure 4 shows a schematic of the digital signal processing.

## 1.a. Two-way Timing Equations

After acquisition, the timing discriminator outputs the time offset between the incoming comb pulse train and the tracking TPFC at 40 kSa/s. This offset is digitally summed with the local tracking TPFC timing to generate time samples of the incoming pulse train, $t_A$ and $t_B$, at each site. The time samples depend on the time offset between the sites' reference oscillators, $\Delta T_{osc}$, the synchronization feedback $\Delta T_{cntrl}$ at site B ($\Delta T_{cntrl} = 0$ for open-loop operation), the reciprocal time-of-flight, $T_{tof}$, and noise terms. Specifically,

$t_B = -(\Delta T_{osc} - \Delta T_{cntrl}) + T_{tof} + \varepsilon_{turb,B} + \varepsilon_{qn,B} + \varepsilon_{comb,B}$ where $\varepsilon_{turb,B}$, $\varepsilon_{qn,B}$ and $\varepsilon_{combs,B}$ are the zero mean added noise terms respectively from turbulence, quantum noise (i.e., photon shot noise), and the system noise floor (from jitter in the frequency comb phase locks and drifts in the transceiver pathlengths). $t_A$ reverses the sign of the first term and exchanges B → A in the subscripts. Their linear combinations+

$$\Delta t = \frac{t_A - t_B}{2} = (\Delta T_{osc} - \Delta T_{cntrl}) + \varepsilon_{NR,turb} + \varepsilon_{qn} + \varepsilon_{combs}$$
$$\bar{t} = \frac{t_A + t_B}{2} = T_{tof} + \varepsilon'_{turb} + \varepsilon'_{qn} + \varepsilon'_{combs}$$
(M1)

yield, respectively, a measure of the relative time offset between the two clock combs and the time-of-flight over the 300-km link, both modulo half the comb pulse period of 2.5 ns in initial acquisition. Once measurements of the time samples commences, both $\Delta t$ and $\bar{t}$ can be tracked across many comb pulse periods. (The equation for $\Delta t$ appears as Eqn. (2) in the main text.) The noise terms are the corresponding asymmetric (unprimed) and symmetric (primed) combinations of the previously defined site-specific noise. The uncorrelated quantum noise will have the same standard deviation for both, given by Eq. (1). This quantum noise will dominate the two-way time offset measurement $\Delta t$ at low received power if the turbulence-induced noise is fully reciprocal, i.e. $\varepsilon_{NR,turb} = 0$.

## 1.b. Timing Discriminator response and calibration

As shown in Supplemental Figure 5, the optical timing discriminator utilizes the birefringence of PM optical fiber to generate two interferogram channels where the lead and lag positions of the incoming clock comb pulse and local tracking comb pulses are switched [29]. The heterodyne signal output from each balanced detector is demodulated by the expected frequency offset between the comb pulses. This frequency offset must be known a priori to ±13 kHz during acquisition, as discussed below, but after acquisition the frequency offset is tracked and adjusted to center the demodulation frequency. (Not shown in Supplemental Figure 5). The raw output of the timing discriminator is then the magnitude of the two demodulated signals, $|V_1|$ and $|V_2|$. After calibration, the sum of their squares provides a measure of the incoming clock comb pulse train, in Watts, and their normalized relative values provides a measure of the relative timing of the incoming clock comb pulse train with respect to the local tracking comb, in femtoseconds.

## 1.b.1 Timing Discriminator: Signal power calibration

As with previous comb-based OTT using linear optical sampling, the input power is measured by calibrating the heterodyne signal amplitude [15,24] but in this case the signals from the timing discriminator are used. (Direct detection of the incoming power at the 100 fW level would otherwise be challenging.) For this heterodyne detection, the power is given by $P = C_p \left[ \left( |V_1|^2 + |V_2|^2 \right) / 2 \right]$, where the calibration factor, $C_p$ has units of $\left[ W/V^2 \right]$ and is a function of the local tracking comb power on the detector, pulse shape and dispersion, and timing imbalance of the optical timing discriminator. To measure this calibration constant, the link is shorted, attenuated to low powers, and synchronized. We then measure the value of $\left( |V_1|^2 + |V_2|^2 \right) / 2$. Next, we measure the total clock comb optical power at the input of the four balanced detector input ports using a calibrated power meter. Based on repeated calibrations, we expect this calibration constant to be accurate to within ±20%.

This power calibration applies to the normal operation where the demodulated signals are approximately half their peak values (see Supplemental Figure 4). During acquisition, the two demodulated squared signals are added together with a temporal delay that reflects the linearly scanned tracking comb. As a result, the peak demodulated voltages are ~2× higher for the same optical power, and the calibration factor is ideally 4× lower. In reality, for the delays here, the increase is closer to a factor 2× since the operation point is not fully half of the peak voltage (see Supplemental Figure 5), giving $C_{p,acq} \approx C_p / 2$.

## 1.b.2 Timing Discriminator: Timing error calibration

The timing error is calculated based on the normalized difference of the two demodulated voltages by $\delta t = f_{td}(E)$, where $E = (|V_1| - |V_2|)/(|V_1| + |V_2|)$ and $f_{td}(E)$ is the calibration function. $f_{td}(E)$ is determined in a separate shorted calibration by deliberately offsetting the clock and tracking comb pulse and measuring the corresponding error signal, as shown in Supplemental Figure 5. Since the error signal is normalized by the sum of the two demodulated signals, the timing error is insensitive to common-mode (e. g. turbulence-induced) amplitude fluctuations between the channels. Furthermore, the performance of the overall system is insensitive to small changes in the calibration function, for example due to changes in the differential pulse dispersion, since the feedback seeks to lock this error signal $\delta t$ to zero (see Supplemental Figure 4). Moreover, the fact that the timing discriminator's differential delay arises from the difference in propagation constant of the two fiber polarization modes, rather than a mechanical path length difference, means that the temperature stability requirements are quite loose.

## 1.b.3 Timing Discriminator: Quantum limited operation

Both quantum-limited acquisition and quantum-limited operation depend on the signal-to-noise ratio (SNR) of the timing discriminator. If that SNR is shot-noise limited, then the timing discriminator is operating at the standard quantum limit. We consider that limit under the assumption of heterodyne detection.

We first consider acquisition. Acquisition uses the average mean-squared voltage detected by time timing discriminator, $(|V_1|^2 + |V_2|^2)/2$ to search for pulse overlap. The detector noise level is 1.7 dB above shot noise set by the local tracking comb. Based on this noise level, we very conservatively set a threshold value of $\sim(98\mu V)^2$, which yields a false trigger rate of less than once per day. Given $C_{p,acq}$, this threshold corresponds to $P_{thresh,acq} \approx 135$ fW, or ~20 photons for the sample time of $(2 \times 26\text{kHz})^{-1} = 19\mu s$ (or 0.005 mean photons per comb pulse) corresponding to the 26-kHz heterodyne detection bandwidth. A lower detection threshold than 135 fW is certainly possible by accepting a higher false detection rate than 1/day. However, since this tradeoff is very steep, only a small decrease in threshold is realizable. In principle, one could also increase the averaging time to reduce the power threshold. However, the averaging time is limited by the turbulence-induced decoherence over the link.

Next, we consider quantum-limited operation of the timing discriminator in its estimate of the timing error, $\delta t$, between the clock comb and tracking comb pulses, e.g. the value of $\gamma$ in Eq. (1). To find the quantum-limit, we consider heterodyne detection between two, unchirped, gaussian pulses offset in time. (The gaussian shape avoids sidelobes that can lead to false timing error values and also corresponds to the gaussian filters used in our apparatus.) This simple configuration is equivalent to considering the output from only one arm of the timing discriminator. For timing measurements, we define $\delta t = 0$ to correspond to a pulse separation equal to the full-width half-maximum, $\tau_p$, of the pulse intensity. As discussed in Ref. [29], this yields a timing error of Eq. (1) with $\gamma_{ql,1} = \left(2\ln(2)\sqrt{\eta}\right)^{-1}$, where $\eta$ is the detector quantum efficiency. For the two-way time transfer, assuming equal number of received photons at each site, there is an additional factor of $\sqrt{2}$, giving

$$\gamma_{ql} = \frac{1}{2\sqrt{2}\ln(2)\sqrt{\eta}} \approx 0.6 \qquad (M2)$$

In the actual system, we use the dual outputs of the unbalanced Mach-Zehnder interferometer. In that case, $n$ is the total number of received photons. In addition, there is a ~1.7-dB penalty, $D$, due to detector noise, as discussed above. Finally, in our system, the largest penalty results from differential dispersion between the incoming clock comb pulses and the local tracking comb pulse. To lowest order, this causes broadening of the temporal signal by a factor $C$ and a corresponding reduction in peak height by $C$, leading to a $C^2$ increase in $\gamma$. The differential dispersion contains higher orders that lead to distortions of the pulse shape beyond broadening, but this first-order model is sufficient to characterize the degradation in the factor $\gamma$, as

$$\gamma = \left(DC^2\right)\gamma_{ql} \qquad (M3)$$

Experimental values for $\gamma$, $C$, and other relevant parameters are given in Supplemental Table 1. We set a power threshold for a "valid" timing discriminator measurement of 264 fW at site A and 283 fW at site B (see Supplemental Figure 2). These are chosen very conservatively to remain in the linear regime of the timing discriminator curve (Supplemental Figure 5) despite the additive shot noise in the detected signal (and including the 1.7 dB detector noise penalty). At this threshold, the additive noise leads to a noise level in $E$ of $\pm 0.3$ (2-sigma), which covers 40% of the shaded region in Supplemental Figure 5. This power level corresponds to ~40 photons in a sample time of $(2 \times 26\text{kHz})^{-1} = 19 \mu s$ (or 0.01 mean photons per comb pulse).

### 1.c. Dispersion balancing (over air)

Since both the optical timing discriminator calibration and signal power calibration were performed over fiber shorted links, it is important that the dispersion of the pulses sent over 300 km of air matches the dispersion of the pulses sent over the shorted link used in the calibrations, both for the timing discriminator and the power. Therefore, we must compensate for the additional group delay dispersion of the 300 km of air. Using 70,000 pascal for air pressure, 10 ℃ for air temperature, 25% relative humidity, and 450 ppm for $CO_2$ mole fraction, we estimate the group velocity dispersion of the air in Hawaii to be $\beta_2 = 7.43 \times 10^{-30} \, \text{s}^2/\text{m}$. In comparison, at 1560nm, PM1550 fiber has dispersion of $\beta_2 = -2.18 \times 10^{-26} \, \text{s}^2/\text{m}$. Therefore, we can compensate for the 297 km of air by adding 101 m of PM1550 fiber to the fiber path length already in place for a shorted measurement. To fine tune the dispersion matching, we measure the pulse widths across the atmosphere and minimize the width of the interference signal, which matches the width used in the fiber-shorted calibrations of the system. This optimization yields a total PM1550 fiber length of 105 m.

## 1.d. Truth data measurement

The folded-link geometry used here allows for out-of-loop verification of the synchronization by measuring the relative timing of the pulses from the clock comb on site A and the synchronized clock on site B. The clock comb light from site A was routed to site B via a fiber and combined with the synchronized clock comb light on site B. This combined light was filtered with a 12-nm-wide filter centered at 1560 nm before detection on a balanced photodetector. The resulting signal is then demodulated to yield a voltage that varies with the relative timing between pulses. (Because there is no scintillation present, we do not use the full timing discriminator approach as with the tracking comb.) This signal peaks when the pulses directly overlap in time and vanishes if the pulses are much more than a pulse-width apart, as shown in Supplemental Figure 6, which was generated by digitally stepping the offset applied to the synchronization feedback loop. For normal operation, we set the synchronization calibration offset so that zero time offset corresponds to the value shown in Supplemental Figure 6. This central location provides the maximum dynamic range for measurements of the out-of-loop timing. From a 5$^{th}$ order polynomial fit to this curve, we generate a calibration function that yields the out-of-loop time offset as a function of measured voltage.

## 1.e. Signal acquisition via the Kalman Filter

At the onset of operation or after a long fade, the system must acquire the tracking lock between the local tracking comb and the incoming clock comb. The first step in the acquisition is to establish the optical link between the two sites with the free-space optical terminals as described below so that the incoming clock comb pulses are coupled into single-mode fibers on both sides and directed to the timing discriminator. The demodulation frequency used in the timing discriminator is then set to the appropriate value, to within the 26 kHz detection bandwidth, given the known locking conditions of the frequency combs and *a priori* information on the relative frequency between sites. This 26 kHz frequency knowledge corresponds to a fractional frequency knowledge of $6.8 \times 10^{-11}$ at the optical carrier frequency. It can be provided by (i) measuring the reference oscillator frequency against a GPS disciplined oscillator, (ii) use of a coarser frequency transfer over a comms link as has been done previously, or (iii) past estimates from a Kalman filter. Alternatively, the acquisition could include a search over frequency, but this additional search has not been implemented here. The demodulation frequency is set to differ between sites, nominally at values of 10 MHz and 12 MHz to avoid cross talk from internal reflections within the free-space optical terminals.

Finally, the system initiates a Kalman-filter based search for temporal overlap between the incoming clock and local tracking comb pulses. As discussed in the main text, this search is based on the timing discriminator signal power output value of $\left(\left|V_1(t)\right|^2 + \left|V_2(t - r\tau_p)\right|^2\right)/2$, where $r$ is the scanning rate of the tracking comb. This signal is proportional to the input clock comb pulse power. A voltage threshold is applied that corresponds to ~135 fW at zero pulse time offset (See 1.b.1).

The search for temporal overlap is complicated by turbulence and by differential clock drift between the unsynchronized sites. Intensity scintillation from turbulence causes the signal intensity to fluctuate randomly, complicating the mapping between peak heterodyne signals and the location of the incoming clock comb pulses in time. Additionally, differential drift between the two cavity-stabilized lasers will cause the incoming clock comb pulses to move in time relative to the local comb pulses on the timescale of the search. To overcome these challenging conditions, we tightly couple the acquisition algorithm with a Kalman filter which keeps track of the estimated position of the incoming clock comb time and most critically, its associated uncertainty.

Starting from a completely unknown received comb timing with respect to the tracking comb, the search controller sweeps the tracking comb through the full 5-ns non-ambiguity range until a signal is observed above threshold, indicating a momentary coincidence between the tracking comb and the received clock comb. This signal is then used to update the Kalman filter's state and uncertainty estimates.

The effect of the Kalman filter is to aggregate these intermittent observations of the temporal overlap into a consistent picture for the estimated trajectories of $t_a^{KF}$ and $\sigma_{KF,ta}^2$ (for the measurement of the received clock comb pulse time from site B at site A). These are in turn used by the search controller to decide the search pattern swept by the tracking comb. Specifically, the tracking comb timing is swept as a triangle waveform through a search space set by $t_a^{KF} \pm 3\sigma_{KF,ta}$. In proper operation, each individual observation of the interference signal improves the accuracy of $t_a^{KF}$ and reduces its estimated uncertainty $\sigma_{KF,ta}^2$, narrowing the size of the search span and rapidly converging on the true position of the received comb pulses. Finally, once the position uncertainty is small enough ($\sigma_{KF,ta} < 2\tau_p$), the tracking comb control switches to tracking $t_a$ instead of having to sweep.

In practice, given the good performance of the cavity-stabilized CW reference lasers and the high timing accuracy of each detection of temporal overlap between the tracking and incoming clock comb pulses, the first observation event dramatically collapses the uncertainty of the prediction from 5 ns to ~1 ps. This then leaves almost exclusively a contribution from the unknown initial frequency offset between the sites, which produces a ramp in time offset, or a linear varying received comb timing.  As soon as a second observation is produced, the frequency offset is determined and the uncertainty becomes very small, dominated by the random link delay fluctuations.  Depending on the received link fades, acquisition either completes almost immediately after these initial observations, or the tracking comb sweeps over a small range centered on the predicted location until the received power crosses threshold again.

However, at the lower received powers, there is a significant probability that the signal is not observable (below threshold) at the moment in the sweep that the tracking comb and incoming clock comb pulses overlap temporally.  This lack of signal does not have a direct effect on the Kalman filter's estimates, but it does have an indirect one: the uncertainty in the clock time offset grows over time in the absence of observations due to the inherent statistics of the reference oscillator noises.  Thus, if too much time passes between observations, the search space will naturally increase, up to the maximum of the full 5-ns search space, essentially resetting the search process.

Similarly, when the system is in tracking mode but experiences a fade, the Kalman filter's uncertainty can be used to decide whether a new search is necessary ($\sigma_{KF,ta} > \tau_p$), and if so, the search can occur over a fraction of the 5-ns non-ambiguity range if $6\sigma_{KF,ta} < 5$ ns , greatly speeding up the process.

**1.f.  Design and operation of the free-space optical terminals**

The terminals are designed to offer low (~ 1 dB) through-terminal optical loss for measuring two-way optical time transfer across a free-space optical link, while providing active tip-tilt correction for turbulence-induced beam motion.  The terminals are fiber coupled, single-transversal-mode, and fully reciprocal for the science (comb) light thus preserving link reciprocity for two-way time transfer measurements.

To avoid sacrificing comb light to active tip/tilt stabilization, a separate beacon beam at 1532 nm or 1542 nm is polarization multiplexed directly onto the comb beam.  A servo system, comprised of angular detection of the received beacon and a tip/tilt galvo mirror, provides tip/tilt correction of the received beacon.  This maintains position of the received comb light on the core of the single-mode fiber, while also providing forward correction for the beams exiting the terminal.

The beacon wavelengths differ for the two terminals, allowing wavelength demultiplexing of the transmitted and received beacon beams.  A terminal's transmitted beacon beam, originating from a polarization maintaining fiber-coupled laser, is collimated through a fiber collimator, reflected off the face of an interference filter (which acts as a mirror), and directed to the polarization coupler. The combined, ~3.3-mm diameter beam is directed to a galvo mirror that provides tip/tilt control, and then passes through a 24:1, 10 cm aperture beam expander. At the receive terminal, the combined beam enters the expander and reflects off the galvo mirror. The beacon is demultiplexed from the received time transfer beam and is directed to the interference filter. The received beacon passes through the filter and is directed to the focal plane array (FPA) of an InGaAs camera. For operation at low received powers, the beacon is synchronously modulated and detected at ½ the camera frame rate, allowing for real-time background subtraction. The beam position on the FPA is maintained at a pre-calibrated location through feedback to the galvo mirror. Additionally, the beacon power integrated over the region of interest is recorded at a ~25 Hz sampling rate.

The use of the galvo mirrors and FPA provides means of a broad search for the beacon return enabling link acquisition after coarse initial pointing. A visible camera with telescopic lens provides the gross initial pointing, aided by an 830-nm, broad angle beacon transmitted through the cat-eye reflector and a matched 830-nm bandpass filter in front of the camera.

**1.g. Estimation of turbulence strength from beacon scintillation measurements**

A variety of statistical models have been developed for turbulence-induced power fluctuations, i.e. scintillation and fading, including the lognormal, Gamma-Gamma [44] and modulated Gamma-Gamma [45]. The Gamma-Gamma is often used because it is physically motivated and reproduces real observed scintillation statistics reasonably well over a wide range of turbulence strengths [44]. However, in the context of long-range laser propagation it suffers from a substantial limitation in that it does not include the effects of large-scale beam wander. A statistical model of beam wander can be integrated into the Gamma-Gamma distribution, and the resulting model is called the modulated Gamma-Gamma [45].

Our free-space optical terminals record the received beacon power at a rate of approximately 25 Hz using the measured sensitivity of the InGaAs camera which records the incoming beacon signal. As illustrated in Supplemental Figure 11, the shape of the resulting irradiance probability distribution is strongly inconsistent with a Gamma-Gamma distribution but is well-fit by a modulated Gamma-Gamma which indicates that beam wander is significant over the 300-km round-trip distance. We time-segment the records of beacon power and perform a maximum-likelihood fit to the observed values of link loss to yield (Supplemental Figure 12) a time-varying estimate of $C_n^2$ and a multiplicative, non-turbulence excess link loss. The excess loss is primarily attributed to a combination of the retroreflection losses and the known defocus of the beacon (compared to the comb light) from chromatic aberration in the 24:1 beam expander.

**REFERENCES (METHODS)**

# SUPPLEMENTAL MATERIAL

## 1. Performance limitations

The timing jitter on the value of $\Delta t$ from Eq. (2) can be expressed in terms of its timing power spectral density (PSD), $S_{\Delta t}$, from which one can calculate both the TDEV and MDEV. From Eq. (2), this PSD is given by

$$S_{\Delta t}(f) = S_{combs}(f) + S_{qn}(f, P) + [1 - H(f)] S_{osc}(f) + S_{NR, turb}(f) + S_{dpiston}(f, \tau_{delay}) \quad (S4)$$

where $S_{combs}$, $S_{qn}$, $S_{osc}$ and $S_{NR,turb}$ are the PSDs of $\varepsilon_{combs}$, $\varepsilon_{qn}$, $\Delta T_{osc}$, and $\varepsilon_{NR,turb}$. $H$ is the effective closed-loop synchronization transfer function. The final added term $S_{dpiston}(f, \tau_{delay})$ is the non-reciprocal portion of the turbulence-induced piston effect due to the delay between the counter-propagating pulses in the turbulence layer. It is negligible for the Hawaii link, but non-negligible for a link to a geosynchronous orbit. As will be seen, most of these terms lead to white noise at Fourier frequencies in the 0.1-100 Hz range that dominate the TDEV and MDEV for averaging times, $\tau$, of 0.01 to 10 s. For white timing noise, $S_x$, the corresponding TDEV and MDEV are[46]

$$TDEV = \sqrt{\frac{S_x}{2\tau}}$$
$$MDEV = \sqrt{\frac{3 S_x}{2\tau^3}} \quad (S5)$$

## 1.a. Comb Noise

At high (nW-level) received powers and low turbulence (or shorted measurements), the performance is limited by the comb timing jitter, $\varepsilon_{combs}$, at short averaging times. The repetition rate of the comb is given by $f_r \equiv (\nu_{cav} - f_{opt} - f_{CEO})/m$ where $\nu_{cav}$ is the frequency of the cavity stabilized laser, $f_{opt}$ is the locked RF frequency offset between the cavity stabilized laser and the nearest comb tooth with mode number $m$, and $f_{ceo}$ is the locked carrier envelope offset frequency. The timing noise is a scaled version of the phase noise on the repetition rate. To find this phase noise, we measure the phase noise from the two locks, $f_{opt}$ and $f_{ceo}$, up to the single-sided measurement bandwidth of 13 kHz. These quantities are assumed to be uncorrelated, so we add them in quadrature and divide by $m$ to find the phase noise on the repetition rate, which is then scaled to timing noise for a given comb. We then add the timing jitter between the two combs being heterodyned (local tracking combs and remote clock combs) again in quadrature to find the timing error. Since this timing error is measured on two channels (Supplemental Figure 3) we divide this timing noise by $\sqrt{2}$. For site A, this timing noise is 220 as and for site B, it is 160 as. Adding these in quadrature gives the noise floor in Figure 2b of 270 as. At very long times (low Fourier frequencies), temperature-induced drifts in the transceivers add noise, which can effectively be included in a generalized $S_{combs}$, which is then interpreted as a system noise floor.

### 1.b. Quantum noise (shot noise)

The shot noise on the timing signal leads to white timing noise as given by Eq. (1). Converting to a PSD gives

$$S_{qn}(f,P) = 2\gamma^2 \tau_p^2 \left(\frac{h\nu_{opt}}{P}\right) \quad (S6)$$

For $\gamma = 3.4\gamma_{ql}$ seen in the 2-km data and using Eq. (S2), this gives $TDEV_{qn} = (246 \text{ as})/\sqrt{P\tau}$ and $MDEV_{qn} = 4.3 \times 10^{-16}/\sqrt{P\tau^3}$, where $P$ is the one-way received power in pW.

### 1.c. Non-reciprocal piston due to time-of-flight delay

The fundamental reciprocity of a single-mode link requires that the pulses probe the identical turbulence structure, i.e. transit it simultaneously. If there is a delay between pulses that approaches the characteristic time for turbulence to change, then the turbulence-induced path delay is no longer reciprocal and the time-of-flight does not fully cancel when evaluating Eq. (2). Instead, there is an added non-reciprocal noise with PSD,

$$S_{dpiston}(f, \tau_{delay}) = 4\sin^2(\pi f \tau_{delay}) S_{piston}(f) \approx (2\pi f \tau_{delay})^2 S_{piston}(f) \quad (S7)$$

where $S_{piston}(f) \propto f^{-8/3}$ is the full piston noise on the time-of-flight. This equation ignores higher order effects for a distributed turbulence layers that do exist for the Hawaii link but are negligible. For the GEO-to-ground link, assuming the time samples are roughly synchronized, $\tau_{delay} = T_{tof}$. Near 1 Hz, $S_{dpiston}(f, \tau_{delay}) \approx (2\pi T_{tof})^2 S_{piston}(f = 1\text{Hz})$. Integration of the Hufnagel-Valley profile with a ground-level turbulence of $C_n^2 = 10^{-14} m^{-2/3}$, at an elevation angle of 45 degrees and from a height of 10 meters, yields $S_{piston}(f = 1 \text{ Hz}) = 17 \text{ fs}^2/\text{Hz}$. From Eq. (S2) and (S4), one finds for a link from GEO to ground, $TDEV_{dpiston} = 2.3 \text{ fs}/\sqrt{\tau}$ and $MDEV_{dpiston} = 3.9 \times 10^{-15} \tau^{-3/2}$, where $\tau$ is the averaging time in seconds.

## 1.d. Non-reciprocal turbulence noise from multi-path interference in strong turbulence

In strong turbulence, short pulses will experience temporal broadening as well as the piston noise discussed earlier. Eqns. 28 and 29 of Ref. [32] gives the following expression for the increased temporal width due to atmospheric turbulence of $\sqrt{T_0^2 + 8\alpha}$ with a broadening parameter $\alpha = 0.391 L C_n^2 L_0^{5/3} / c^2$, where $T_0$ is the transform-limited pulse width, $L_0$ is the outer scale in m, and c is the speed of light in m/s. This broadening parameter, $\alpha$, represents the average increase in the pulse width. We assume here that it sets the scale for the pulse-to-pulse fluctuations in timing, although a more complete theory is needed to describe the added noise in the measured timing. There is no inherent non-reciprocal mechanism for this turbulence effect. However, as the specific pulse parameters (chirp and shape) of both the clock comb pulses and the local tracking comb pulses are not identical at both sites, one might expect only partial reciprocity in the actual two-way measurement. We indeed observe only ~10 dB of suppression (note that the piston noise does suppress to much more than 10 dB; the reduced 10-dB suppression only applies to this pulse broadening noise, which has a much shallower PSD). With this suppression, we estimate an excess timing noise of $\delta t_{excess} = \sqrt{8\alpha/10} \sim 0.56 L_0^{5/6} c^{-1} \left( LC_n^2 \right)^{-1/2}$. For the inset of Figure 4, we assume an outer scale of $L_0 = 1$ m to compare this estimate to the measured $\delta t_{excess}$ of the 300-km Hawaii link. In terms of a PSD, we estimate $S_{NR,turb}(f) \sim 8\alpha/10 f_G$ where the corner frequency is set by the Greenwood frequency, $f_G$ (proportional to the Fried parameter divided by the wind velocity). As noted in the text, for the Hawaii link, this noise PSD is observed to dominate leading to the measured $TDEV = 1.6 \text{ fs}/\sqrt{\tau}$ and $MDEV = 2.8 \times 10^{-15} \tau^{-3/2}$ of Figure 3.

## 1.e. Oscillator noise

Due to the use of a two-way clock comparison, the propagation delay $T_{tof}$ is experienced by both the comb pulses going from site A to site B and again when site B transmits the resulting data back to site A. Thus, the clock feedback loop must work with a system that involves a round-trip delay of $T_{roundtrip} = 2 T_{tof}$, which limits the maximum closed-loop bandwidth. In order to maintain stability, a feedback loop must have positive phase margin at the loop's unity gain frequency. For a system with delay only and an integrator controller, the phase margin in radians is $\pi - 2\pi f_c T - \pi/2$, where $f_c$ is the loop's unity gain frequency, $T$ is the total delay, and the last $\pi/2$ term is the contribution from the integrator. Thus, the maximum stable bandwidth for the closed loop synchronization transfer function, $H(f)$, is strictly less than $f_c = 1/(8 T_{tof})$. The contribution of the term $[1 - H(f)] S_{osc}(f)$ in Eq. (S1) then depends on the oscillator. For the cavity-stabilized lasers used here, $S_{osc}(f) = 7.9 \times 10^{-30} \ f^{-3} s^2 / Hz$ and this contribution is below that of the multi-path noise for the Hawaii link or differential piston noise for the ground-to-GEO link.

## 2. Link Budget

### 2.a. 300-km round-trip Hawaii link

We assume a $1/e^2$ beam diameter equal to 80% of the telescope aperture. We assume Gaussian beam propagation and ignore the effects of clipping of the outgoing beam. The diffraction-limited loss across a link with transmit aperture area, $A_{xmit}$, and receive aperture area, $A_{rcv}$ is then

$$\text{Loss} \approx \ell_{xmit} \ell_{rcv} \ell_{channel} \left[ \frac{A_{xmit} A_{rcv}}{\lambda^2 L^2} \right] \tag{S8}$$

where $\lambda = 1560$ nm is the wavelength, the term in square brackets is the aperture-to-aperture loss due to diffraction, $\ell_{xmit}$ is the extra loss in the transmit system, $\ell_{rcv}$ is the extra loss in the receiver and $\ell_{channel}$ is channel loss due to absorption and scattering. Both $\ell_{xmit}$ and $\ell_{rcv}$ depend on where in the system the power is evaluated. Here $\ell_{xmit}$ is measured from the output of the Gaussian filter for the frequency comb up to the free-space terminal aperture while $\ell_{rcv}$ is measured from the free-space terminal aperture to the balanced photodetectors at the output of the timing discriminator. (See Supplemental Figure 2)

The above expression ignores turbulence. To lowest order for a single mode link across a horizontal turbulent path, we can replace the aperture diameter with the Fried parameter, $\rho$, with a corresponding effective area $A_\rho = \pi (\rho/2)^2$. Based on the measured piston strength and a wind velocity of $V = 10$ m/s, the Fried parameter is $\rho \approx 1.3$ cm so that $A_\rho \approx 1.3 \times 10^{-4}$ m$^2$. Finally, we need to include the cateye reflector, which is located at the mid-point of the link at a distance of $L/2$. In the simplest model, the cateye causes an additional loss $4\ell_{cateye} A_{cateye}^2 / (\lambda^2 L^2) \approx 4.5$ dB where the cateye aperture area is $A_{cateye} = \pi (0.25 \text{ m})^2$ and its transmission loss $\ell_{cateye} \approx 0.5$. The actual loss incurred by the cateye is higher since this simple model ignores the complicated interplay between the cateye aperture, turbulence and diffraction. However, with this simple model, the total loss is

$$\text{Loss} \approx 4\ell_{smf}\ell_{xmit}\ell_{rcv}\ell_{channel}\ell_{cateye}\left[\frac{A_\rho^2 A_{cateye}^2}{\lambda^4 L^4}\right] \quad (S9)$$

where we add an additional loss $\ell_{smf}$ for coupling into the single-mode fiber that is above and beyond the effective aperture reduction. Note the loss scales as $L^4$ rather than $L^2$ because of the cateye. Here, the combined loss transceiver loss, $\ell_{xmit}\ell_{rcv}$, equals -6 dB (see Supplemental Figure 2), and assuming a channel loss estimate of 4 dB ($\ell_{channel} = 0.4$), the values given above give the $\text{Loss}(\text{dB}) \approx 85\,\text{dB} - 10\log(\ell_{smf})$ for typical turbulence conditions. The measured loss is ~91 dB (see Supplemental Figure 2), which implies an additional ~6 dB penalty for coupling into single mode fiber ($\ell_{smf} \approx 0.25$). It is likely at least some of this additional 6 dB penalty is attributable to the use of the cateye reflector. For example, slow beam wander causes the transmit beam to intermittently "miss" the cateye reflector (See Methods 1.g). Finally, turbulence varies and with it the link loss as can be seen in Supplemental Figures 9-10.

## 2.b Long distance links from ground to GEO and further

For geosynchronous orbit to ground, the single mode link loss will again be reciprocal. However, in this case, the turbulence is located very near the ground aperture. Furthermore, the integrated turbulence is much lower; for the Hufnagel-Valley profile at 45-degree elevation, a ground-level turbulence of $C_n^2 = 1\times 10^{-14}\,\text{m}^{-2/3}$ and starting from an elevation of 10 m, the path-averaged $C_n^2 L \approx 2.0\times 10^{-12}\,\text{m}^{1/3}$, which yields a Fried parameter of $\rho = 21\,\text{cm}$. As with optical communication terminals, a reasonable aperture for the space-based terminal is 10 cm. Here, we assume an identical 10 cm aperture on the ground, which is below the Fried parameter and avoids the need for adaptive optics. The estimated loss is then simply,

$$\text{Loss} \approx \ell_{smf}\ell_{xmit}\ell_{rcv}\ell_{channel}\left[\frac{A_{10cm}^2}{\lambda^2 L^2}\right] \quad (S10)$$

where now $L = 35{,}786$ km for GEO. If we assume a reduced channel loss of 2 dB and otherwise assume the same values for the other loss pre-factors as in the current system, the loss is 91 dB, equivalent to the Hawaii link and 11 dB less than the tolerable loss of the system (Supplemental Figure 2). Furthermore, this 11-dB margin is reached with only a 10-cm aperture ground-based telescope located at approximately sea level. Finally, as cislunar orbits are 10× further than GEO, they encounter an additional 20-dB loss. By employing adaptive optics and increasing the ground terminal to a conservative 40 cm, the estimated loss to cis-lunar orbits is reduced to 100 dB putting it in range for the current system.

# REFERENCES (SUPPLEMENTARY MATERIAL)

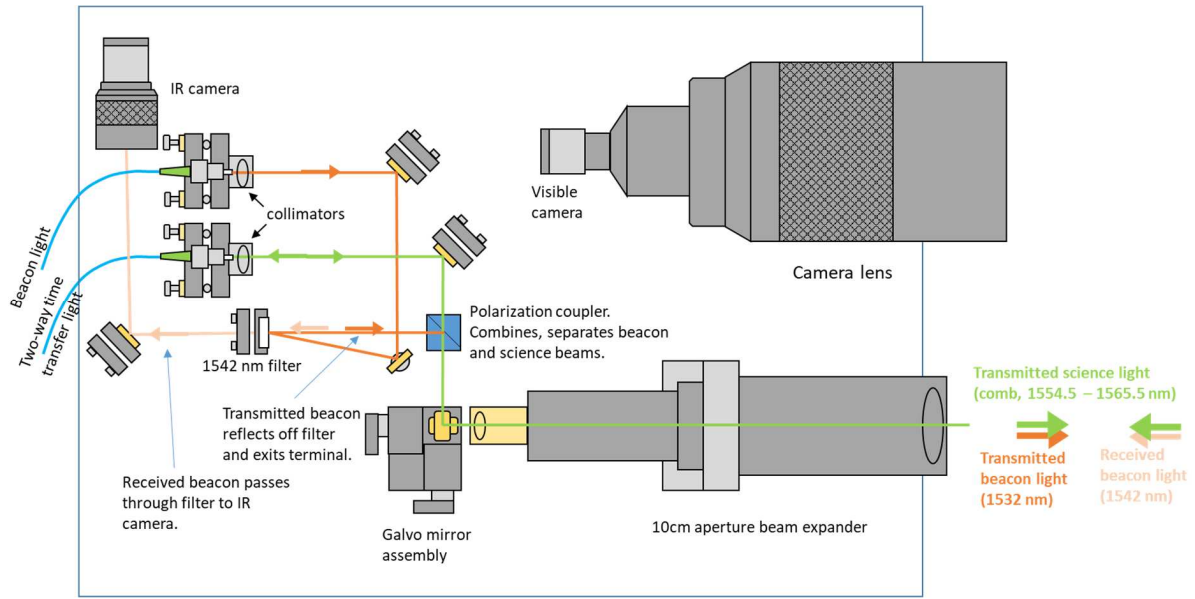

**Supplemental Figure 1:** Free space optical terminal design. Each terminal transmits both the comb light (at a few mW) and a beacon laser signal at similar low power (~2 mW at the aperture) through a 10 cm aperture with an 8 cm $1/e^2$ beam diameter. The beacon light is detected at the far end by a focal plane array. The images are processed, and the beam position fed back to adjust the tip/tilt of the outgoing combined comb light and beacon light. This corrects for atmospheric turbulence and optimizes coupling of the incoming frequency comb light into the polarization maintaining single-mode optical fiber.

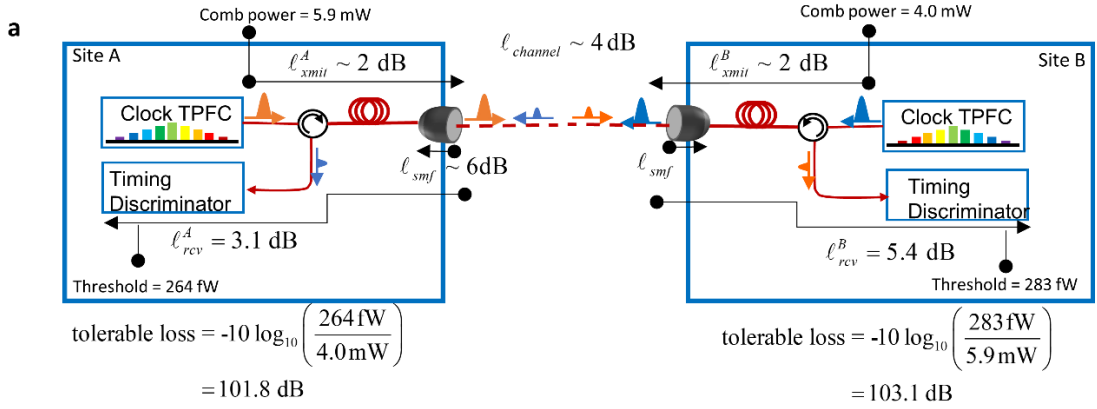

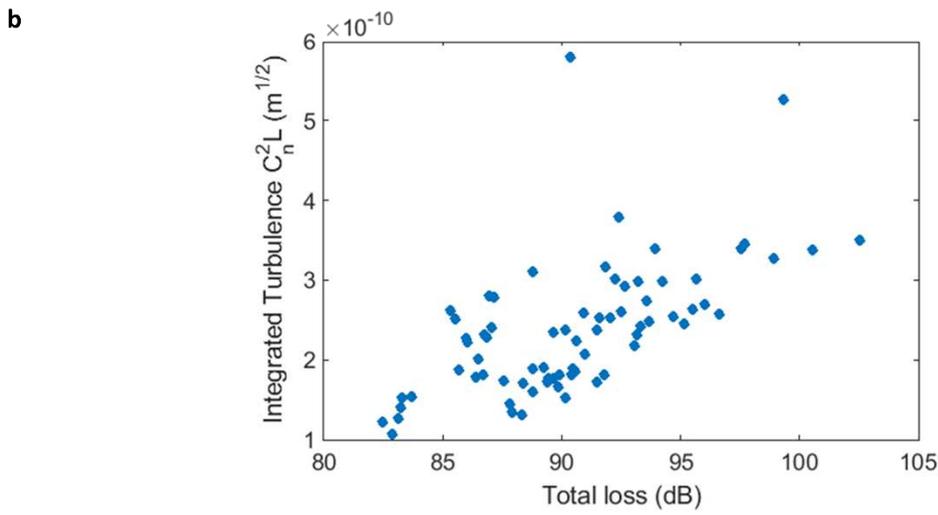

**Supplemental Figure 2:** Total link loss. (a) Diagram of link loss. The aperture-to-aperture link loss, which includes the channel loss $\ell_{channel}$, is reciprocal as is any excess loss from coupling into single-mode fiber, $\ell_{smf}$, whereas the transceiver-specific losses $\ell_{rcv}$ and $\ell_{xmit}$ are not. While $\ell_{smf}$ is reciprocal for both sites, the loss is unidirectional, that is it only occurs when the incoming light is coupled into the fiber. (b) Measured median total loss over different runs plotted against the integrated turbulence, as measured by the piston noise and assuming a wind velocity of 10 m/s. The mean loss across runs is 91 dB giving 11-dB margin over the tolerable loss of 102 dB.

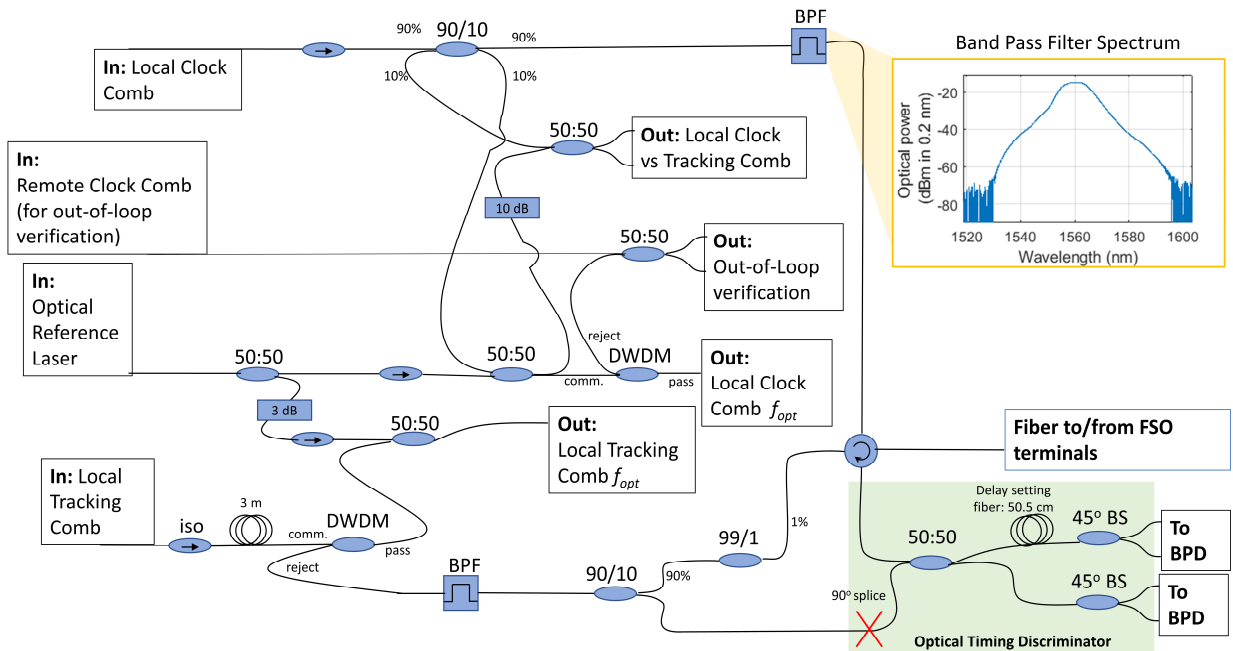

**Supplemental Figure 3**: Optical time transfer transceiver design for a single site. This transceiver routes the local clock comb to the FSO terminals and the incoming remote clock comb to the optical timing discriminator for mixing and detection with the local tracking comb. To minimize excess fiber optic delays, the transceiver also includes the necessary fiber optic components to generate the RF optical beat signals, $f_{opt}$, used for locking the two clock and tracking combs to the local cavity stabilized laser (CW in) reference, and the out-of-loop verification beat signals. A calibration step with a fiber-shorted link determines the time offset for the two clock combs due to path delays in the transceivers. This time offset is included in the overall synchronization loop so that the clock pulses remain overlapped when the system is operated over the link. All fiber is PM1550. 50:50, 50/50 splitter; 90:10, 90/10 splitter; BPF, band-pass filter; iso, isolator; DWDM, dense wavelength division multiplexer at the cavity-stabilized laser wavelength; 45˚ BS, polarization beam splitter with the input fiber rotated 45˚.

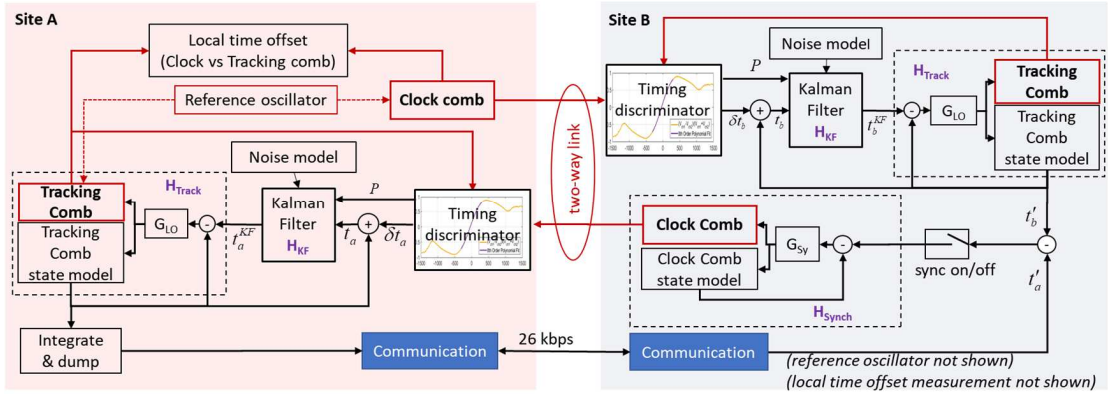

**Supplemental Figure 4**: System diagram emphasizing signal processing. Each site has a local reference oscillator (cavity stabilized laser), to which both the clock comb and tracking comb are self-referenced. As shown for site A, after both the clock and tracking combs are phase-locked to the reference oscillator, they have an arbitrary local time offset between each other. This local time offset is directly zeroed by digitally adjusting the tracking comb timing until there is a maximum heterodyne overlap with the clock comb pulses. Subsequent timing values for the local tracking comb are then referenced to the local clock comb. After acquisition, the timing discriminator measures the (small) time offset $\delta t_a$ between the incoming clock comb and local tracking comb. This timing difference is summed with the tracking comb time offset, to generate an estimate of the incoming clock comb pulse time, $t_a$. This estimate, along with the corresponding measured incoming power, is input to the Kalman filter, whose output provides an optimized, filtered estimate of the incoming clock comb pulse time, $t_a^{KF}$. This value is fed into a final feedback controller for the tracking comb, $G_{LO}$. Due to the feedback loop, the tracking comb output itself (both the physical optical pulse time and the corresponding digital value), is now a filtered, estimated value of the incoming clock comb pulse timing with respect to the local clock comb pulse time. This value is transmitted over a communication link from site A to site B, where it is combined with the corresponding local value of $t_b$ to generate an error signal for the site B clock comb. When both sites have acquired their lock onto the incoming clock comb pulses, the final synchronization feedback controller $G_{sy}$ is activated and the site B clock comb is actively synchronized to the site A clock comb, at their local reference point. The effective bandwidth of the Kalman Filter depends on the input power, and ranges from 10 Hz to 25 Hz. The bandwidth of the subsequent lock of the tracking comb ($H_{track}$) is ~450 Hz. The bandwidth of the synchronization lock ($H_{synch}$) is ~15 Hz. The communication link is over rf coaxial cable here. For a future point-to-point link, it would be over free space by either rf or optical as in Ref. [15]. The message rate (for update of $t_a$) is 400 Hz and the total bit rate is 26 kbps. red solid lines: optical comb pulses, red dashed line: CW optical laser light, black lines: digital values, $P$: input optical power.

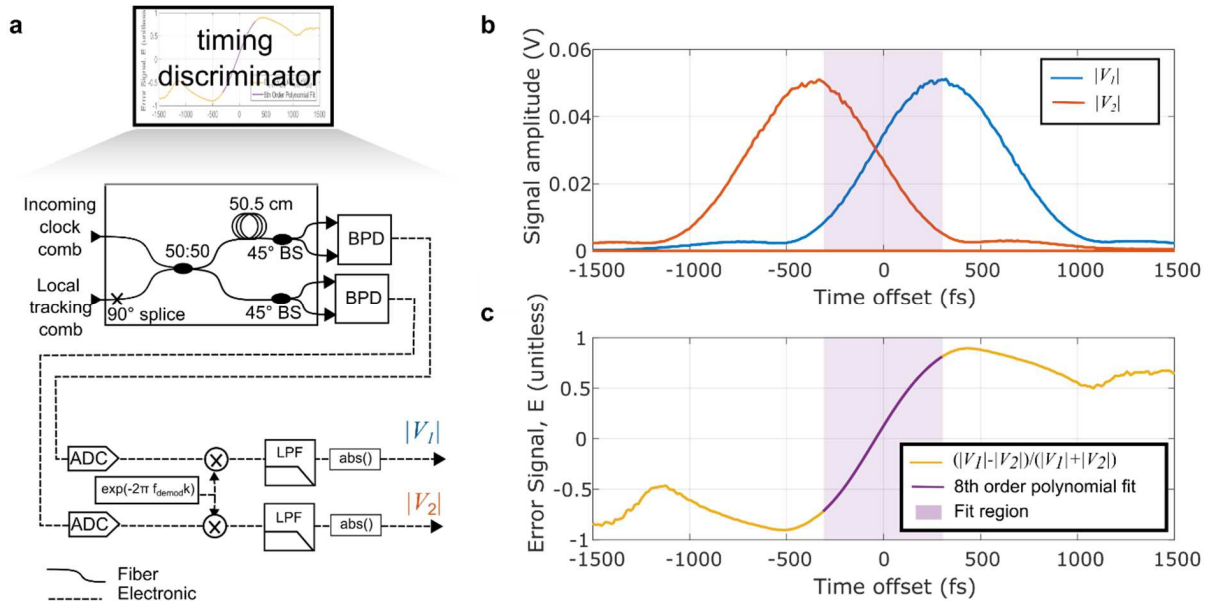

**Supplemental Figure 5**: Optical timing discriminator. (a) System diagram of the timing discriminator. The heterodyne output voltages are demodulated to generate IQ (complex) signals, which are then low pass filtered. The phase of the channel 1 signal is then used in a phase-locked loop (not shown) to adjust the demodulation frequency and center the baseband signals at DC. The magnitudes of the signals are combined to generate an estimate of the power (from their mean squared values) and a timing offset value from their normalized difference. (b) Absolute value of the timing discriminator output voltages, $|V_1|$ and $|V_2|$, as a function of the time offset between the local tracking comb and incoming clock comb. (c) The error signal generated from the two channels along with the polynomial fit used in the digital processing to generate the timing error value based on the normalized error signal, $E$.

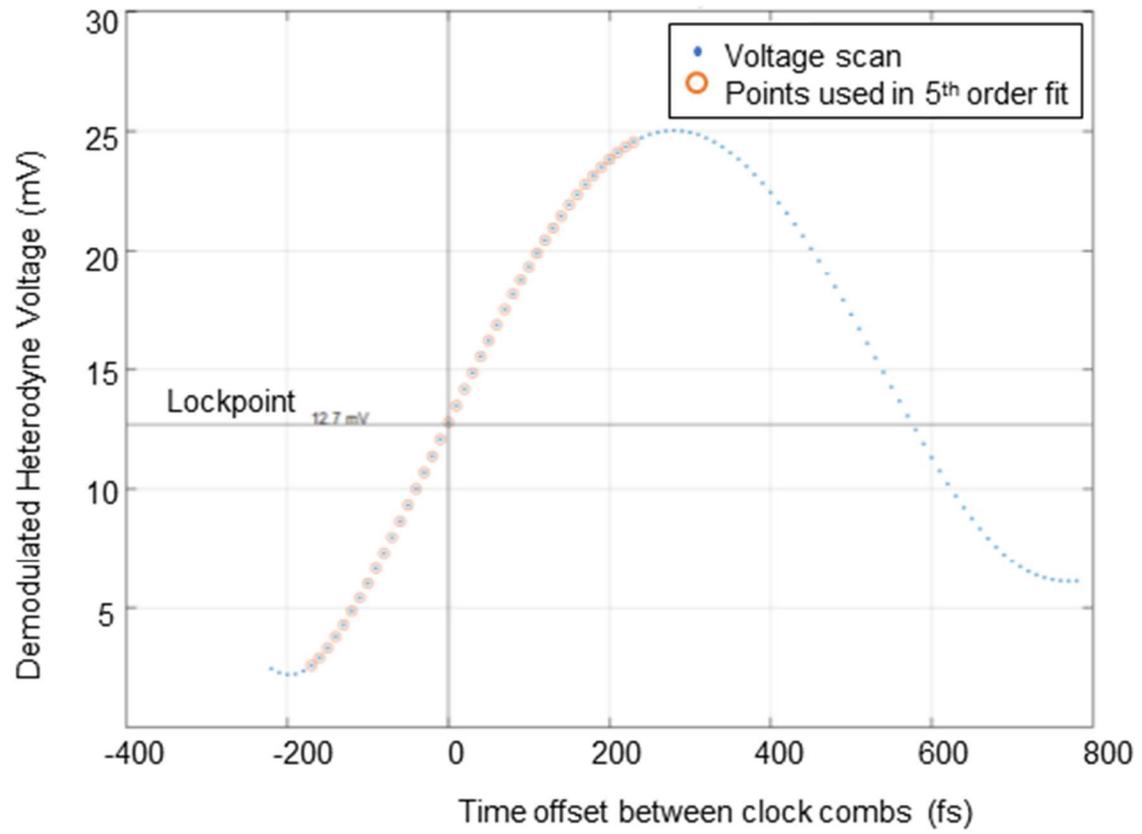

**Supplemental Figure 6:** Out-of-loop verification. Calibration curve of heterodyne voltage vs time offset between the two clock combs used for out-of-loop verification.

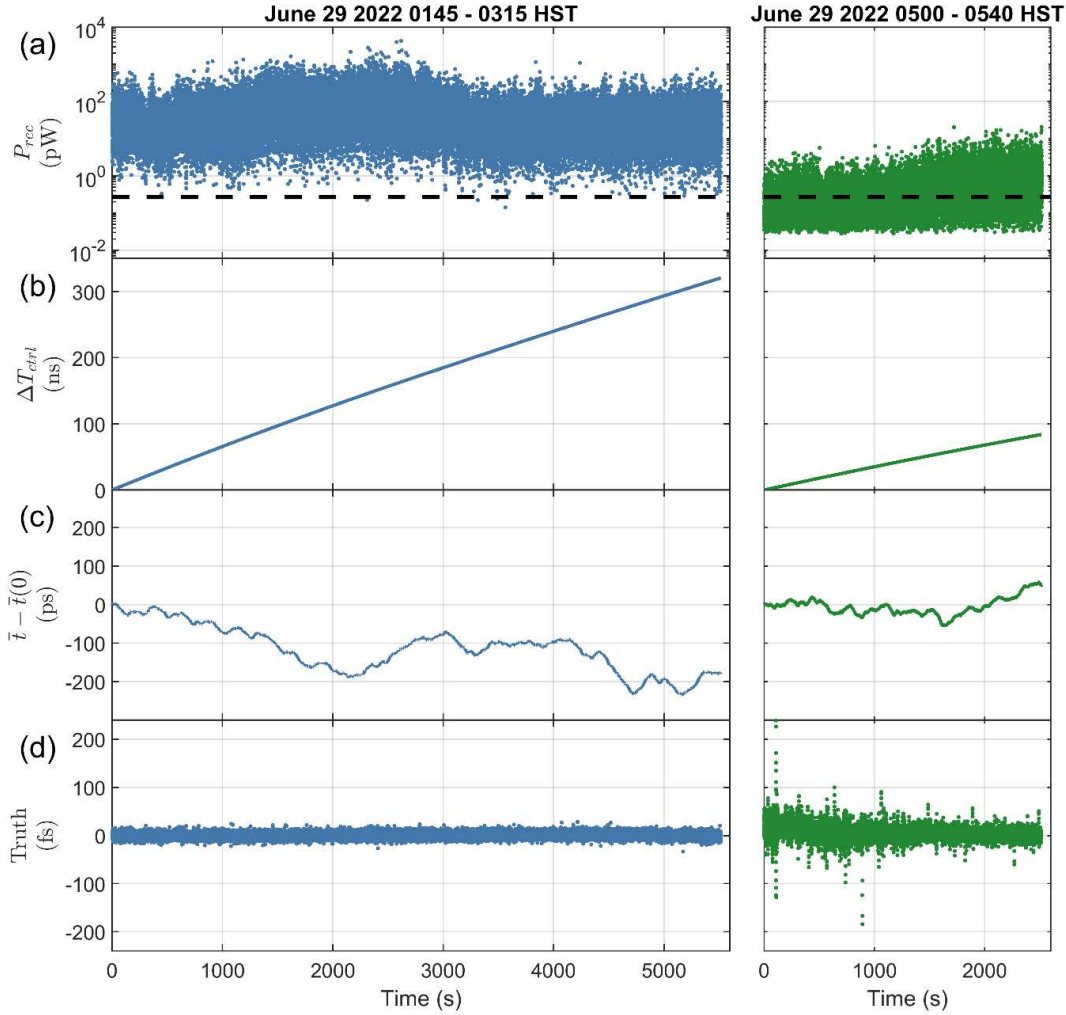

**Supplemental Figure 7:** Traces for synchronization over 300 km. Left are for 4.0 mW comb power at site B (blue curves) and right are from reduced, 40 µW comb power at site B (green curves). (a) Time trace of received power, $P_{rec}$, measured at the output of the timing discriminator for site A with the applied threshold shown as a dashed black line. (b) The control effort, $\Delta T_{cntrl}$, applied on site B to maintain synchronization between the two site's clock combs. As such, it is also a measurement of the time offset between the two cavity-stabilized reference lasers. (c) Changes in the time-of-flight. These fluctuations are due to temperature drifts in the 300-km of air and in the fiber paths up to the terminals, atmospheric turbulence, and mechanical movement in the terminals. (d) The out-of-loop timing verification or 'Truth' data indicates constant temporal overlap between the clock combs at both sites despite 100's of ps changes in the time-of-flight and 100's of ns changes in $\Delta T_{cntrl}$. This truth data is used to generate the instability deviations of Figure 3.

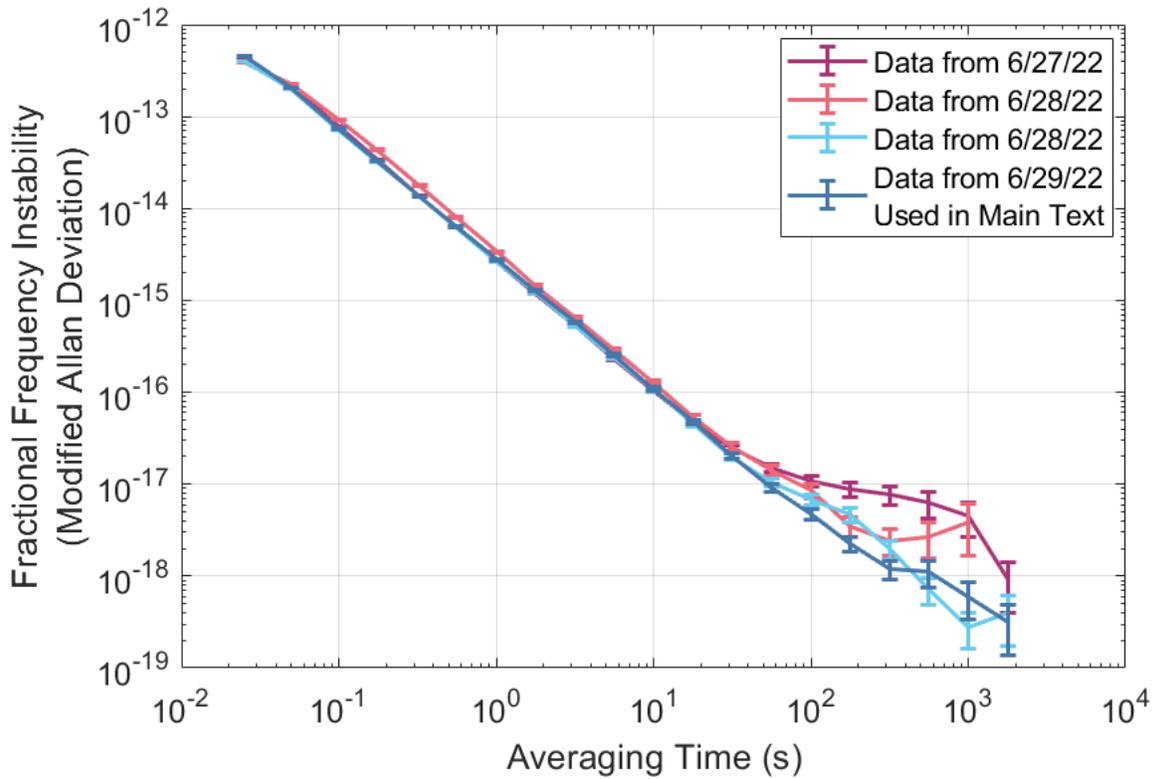

**Supplemental Figure 8:** Additional fractional frequency instabilities. Modified Allan deviations of the out-of-loop verification data ('truth' data) for 4 different days across the 300-km link. Data from June 27th and early on June 28th (purple and pink curves) have an elevated MDEV at long averaging times due to a malfunctioning temperature controller on the out-of-loop verification fiber and both transceivers. The data from June 29th (blue curve) also appears in Figure 3 of the main text.

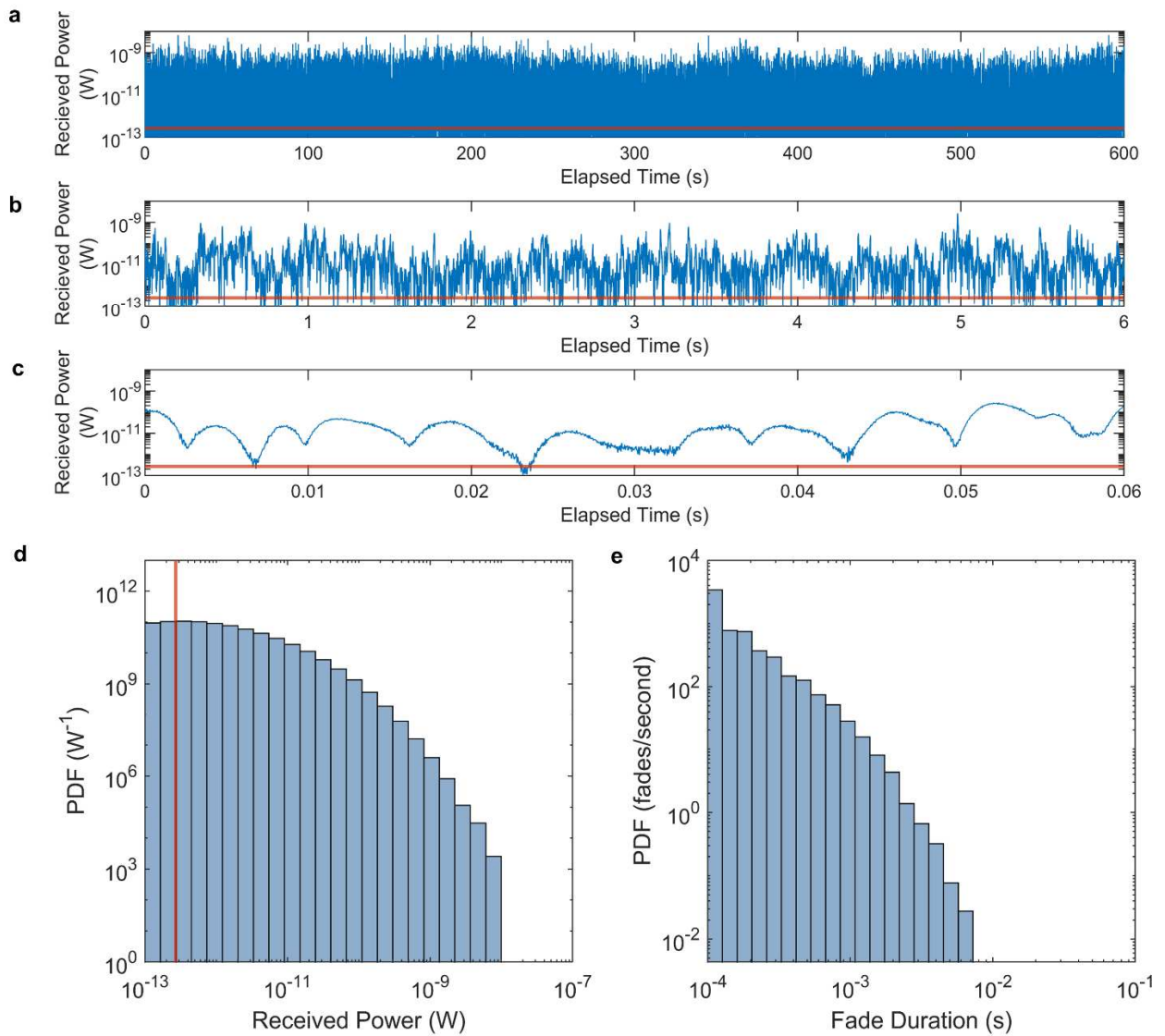

**Supplemental Figure 9**: Received power and fade statistics over the 300 km link at site A for 4.0 mW of comb power from site B.   (a-c) Received power over 600-seconds, 6 seconds, and 0.06 seconds.  Red line indicates the 270 fW detection threshold. (d) Normalized histogram (i.e., probability density function, PDF) of the received power for the 600-second segment. (e) PDF of the fade durations. For the 4.0 mW comb power sent from site B, no fades exceeded a duration of 10 ms – a direct consequence of the low detection threshold at site A.

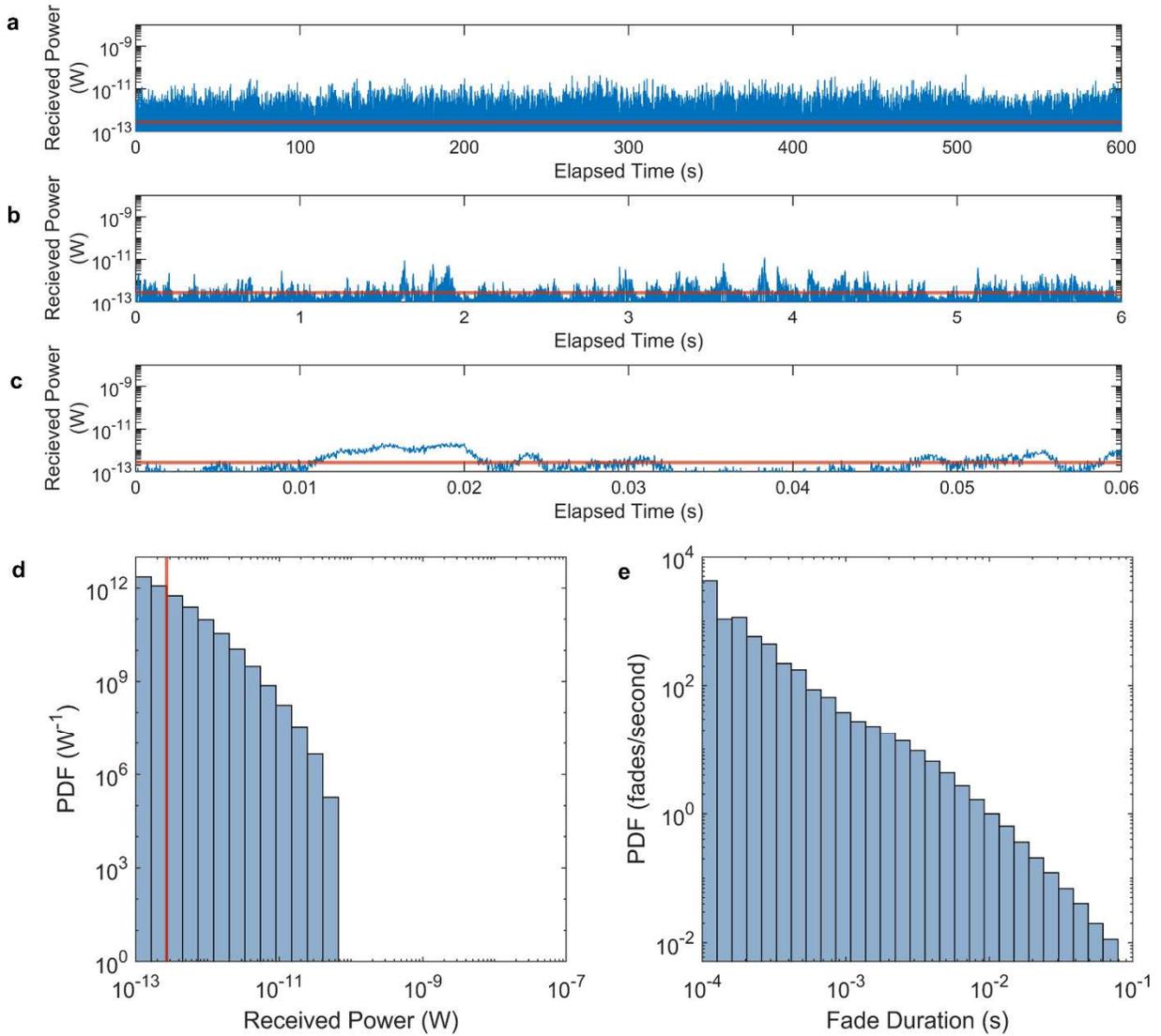

**Supplemental Figure 10:** Received power and fade statistics over the 300 km link at site A for 40 µW of comb power from site B. (a-c) Received power over 600-seconds, 6 seconds, and 0.06 seconds. Red line indicates the 270 fW detection threshold. (d) Normalized histogram (i.e., probability density function, PDF) of the received power for the 600-second segment. (e) PDF of the fade durations. For the greatly reduced launch power, fades are more frequency and of longer duration than the data of Supplemental Figure 9.

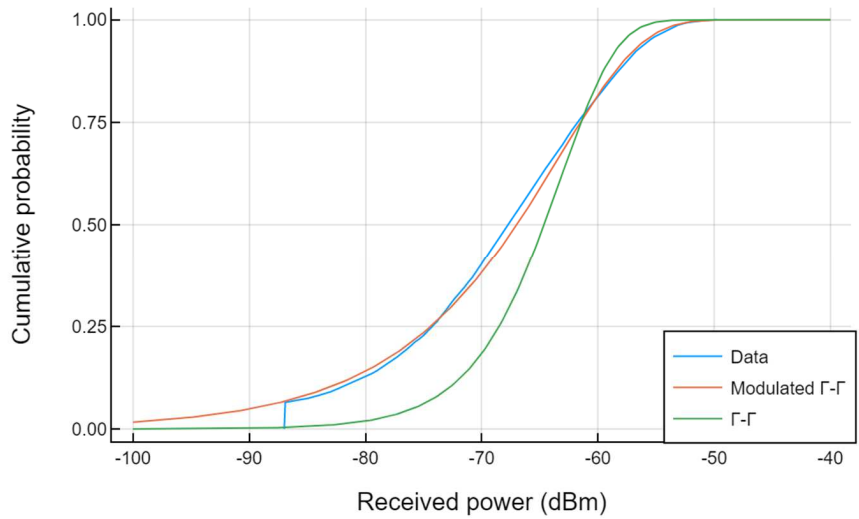

**Supplemental Figure 11:** Comparison of Gamma-Gamma and modulated Gamma-Gamma scintillation probabilities using received beacon power over the 300-km Hawaii free-space optical link. Each fit is a maximum-likelihood estimates two free parameters: $C_n^2$ and a multiplicative excess loss term. The modulated Gamma-Gamma functional form achieves a much better fit to the data, indicating the importance of large-scale beam wander to the observed channel fading statistics.

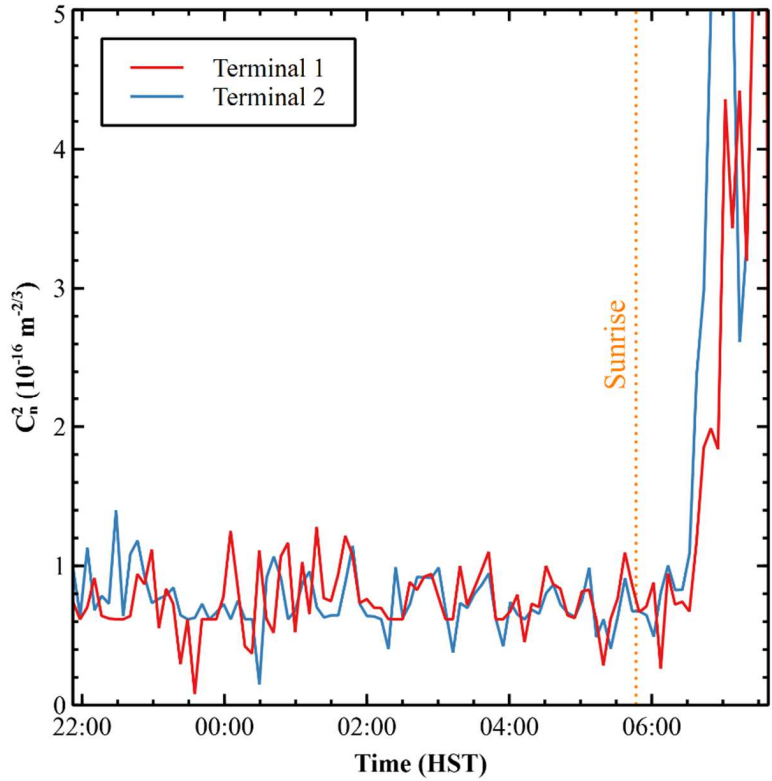

**Supplemental Figure 12:** $C_n^2$ estimation from beacon scintillation measurements. Maximum-likelihood 6-minute estimates of $C_n^2$ from fitting beacon scintillation measurements to a modulated Gamma-Gamma irradiance distribution and a multiplicative excess loss term to correct for retro-reflection. Data collected on 27 June 2022.

**Supplemental Table 1:** Experimental parameters for the data in Figures 2-4 related to quantum-limited operation. Pulse width is defined here as the full width half maximum value. The values for $C$ vary between sites, but the effective averaged value is provided here.

| Quantity | Figure 2 (square filters) | Figures 3 and 4 (Gaussian filters) |
| --- | --- | --- |
| Pulse width for zero chirp, $\tau_p$ | 245 fs | 355 fs |
| Detector quantum efficiency, $\eta$ | 0.80 | 0.80 |
| Detector noise penalty, $D$ | 1.2 (1.7 dB) | 1.2 (1.7 dB) |
| Pulse broadening, $C$ | 1.35 | 1.7 (2-km link) 1.5 (300-km link) |
| Quantum-limit, $\gamma_{ql}$ | 0.6 | 0.6 |
| Actual $\gamma$ | $2.1\gamma_{ql}$ | $3.5\gamma_{ql}$ (2-km link) $2.7\gamma_{ql}$ (300-km link) |